\documentclass[10pt]{article}
\usepackage[utf8]{inputenc}
 \pdfoutput=1

\usepackage{tikz, adjustbox, float} 
\usepackage[authoryear]{natbib} 
\usepackage{amsmath, amsfonts, amssymb, amsthm, dsfont} 
\usepackage{mathrsfs} 
\usepackage{soul} 
\usepackage{bm} 
\usepackage{booktabs}  
\usepackage{enumitem} 
\usepackage{mathtools}
\usepackage{mathabx}

\usepackage{float}

\usepackage{titlesec}
\usepackage{cancel}
\titleformat{\paragraph}
{\normalfont\normalsize\scshape}{\theparagraph}{1em}{}
\titlespacing*{\paragraph}
{0pt}{3.25ex plus 1ex minus .2ex}{1.5ex plus .2ex}

\def\real{{\rm I\!R}}

\DeclareMathOperator*{\argmin}{argmin}

\def\0{{\bf 0}}

\def\x{{\bf x}}
\def\I{{\bf I}}
\def\Y{\mathbf{Y}}

\def\A{{\bf A}}
\def\B{{\bf B}}

\DeclareMathOperator*{\var}{var}  
 
\DeclareMathOperator*{\argmax}{argmax}

\usepackage{latexsym}
\usepackage{amsthm}
\usepackage{amsmath,amssymb,amsfonts,bm,amsbsy,bbm}
\usepackage{epsfig}
\usepackage{graphicx,rotating,lscape}
\usepackage[colorlinks=true,  urlcolor=blue, citecolor=blue, psdextra, pdfencoding=auto]{hyperref}
\usepackage{verbatim}
\usepackage{multirow,color}
\usepackage{geometry}
\usepackage{setspace}
\usepackage{subfigure}
\usepackage{bbm}
\usepackage{ulem} 

\def\x{{\bf x}}

\def\bt{{\boldsymbol\theta}}

\def\0{{\bf 0}}
\def\trans{^{\rm T}}

\def\wh{\widehat}
\def\wt{\widetilde}

\def\var{\hbox{var}}

\def\bse{\begin{eqnarray*}}
\def\ese{\end{eqnarray*}}
\def\be{\begin{eqnarray}}
\def\ee{\end{eqnarray}}
\def\bsq{\begin{equation*}}
\def\esq{\end{equation*}}
\def\bq{\begin{equation}}
\def\eq{\end{equation}}

\def\boxit#1{\vbox{\hrule\hbox{\vrule\kern6pt  \vbox{\kern6pt#1\kern6pt}\kern6pt\vrule}\hrule}}
\def\bse{\begin{eqnarray*}}
\def\ese{\end{eqnarray*}}
\def\be{\begin{eqnarray}}
\def\ee{\end{eqnarray}}
\def\bsq{\begin{equation*}}
\def\esq{\end{equation*}}
\def\bq{\begin{equation}}
\def\eq{\end{equation}}

\def\var{\hbox{var}}

\def\wh{\widehat}
\def\wt{\widetilde}

\def\argmin{\mbox{argmin}}
\def\argmax{\mbox{argmax}}

\def\trans{^{\rm T}}

\def\bg{{\boldsymbol\gamma}}

\def\A{{\bf A}}

\def\B{{\bf B}}

\def\I{{\bf I}}

\def\x{{\bf x}}

\def\Y{{\bf Y}}

\def\log{\hbox{log}}

\def\squarebox#1{\hbox to #1{\hfill\vbox to #1{\vfill}}}

\def\0{{\bf 0}}

\def\var{\hbox{var}}

\def\wh{\widehat}
\def\wt{\widetilde}

\def\log{\hbox{log}}

\newtheoremstyle{mytheoremstyle} 
    {0.3cm}                      
    {0cm}                        
    {\itshape}                   
    {}                           
    {\bf}                   
    {: }                          
    {0em}                       
    {}  

\theoremstyle{mytheoremstyle}

\newtheorem*{Lemma*}{Lemma}

\newtheoremstyle{myExampleRemarkstyle} 
    {0.3cm}                    
    {0cm}                           
    {\itshape}                   
    {}                           
    {\bf}                   
    {: }                          
    {0em}                       
    {}  

\theoremstyle{myExampleRemarkstyle}

\newtheorem{innercustomgeneric}{\customgenericname}
\providecommand{\customgenericname}{}
\newcommand{\newcustomtheorem}[2]{%
  \newenvironment{#1}[1]
  {%
   \renewcommand\customgenericname{#2}%
   \renewcommand\theinnercustomgeneric{##1}%
   \innercustomgeneric
  }
  {\endinnercustomgeneric}
}

\newcustomtheorem{customthm}{Theorem}
\newcustomtheorem{customlemma}{Lemma}
\usepackage{mathalfa}
\usepackage{oubraces}

\let\refBKP\ref
\renewcommand{\ref}[1]{{\upshape\refBKP{#1}}}

\usepackage{titlesec}
\titleformat{\section}{\normalfont\Large\scshape}{\thesection.}{1em}{}
\titleformat{\subsection}{\normalfont\large\scshape}{\thesubsection.}{1em}{}
\titleformat{\subsubsection}{\normalfont\scshape}{\thesubsubsection.}{1em}{}

\titlespacing*{\section}{0pt}{1em}{0em}
\titlespacing*{\subsection}{0pt}{1em}{0em}
\titlespacing*{\subsubsection}{0pt}{1em}{0em}

\usepackage[numbib]{tocbibind}

\usepackage{titlesec}

\titleformat*{\section}{\large \scshape}

\begin{document}

\begin{center}
{	\centering

\large{\textsc{The Generalized Method of Wavelet Moments with Exogenous Inputs: a Fast Approach for the Analysis of GNSS Position Time Series}}

	\vspace{0.5cm} \normalsize
	{\textsc{Davide~A.~Cucci}$^{1}$, \textsc{Lionel~Voirol}$^{1}$, \textsc{Ga\"el Kermarrec}$^{2}$,\\ \textsc{Jean-Philippe Montillet}$^{3,4}$ \& \textsc{St\'ephane~Guerrier}$^{1,5}$}\\
	\vspace{0.75cm} 
	
	{\footnotesize $^{1}$Geneva School of Economics and Management, University of Geneva, Switzerland; $^{2}$Institute for Meteorology and Climatology, Leibniz University Hannover, Germany; $^{3}${Institute Dom Luiz (IDL), University of Beira Interior, Portugal}; $^{4}${Physikalisch-Meteorologisches Observatorium Davos/World Radiation Center (PMOD/WRC), Davos, Switzerland}; $^{5}$Faculty of Science, University of Geneva, Switzerland (e-mail: Stephane.Guerrier@unige.ch).}
	}
	
\end{center}

\begin{abstract}
\noindent  The Global Navigation Satellite System (GNSS) daily position time series are often described as the sum of stochastic processes and geophysical signals which allow studying global and local geodynamical effects such as plate tectonics, earthquakes, or ground water variations. In this work we propose to extend the Generalized Method of Wavelet Moments~(GMWM) to estimate the parameters of linear models with correlated residuals. This statistical inferential framework is applied to GNSS daily position time series data to jointly estimate functional (geophysical) as well as stochastic noise models. Our method is called GMWMX, with X standing for eXogeneous variables: it is semi-parametric, computationally efficient and scalable. Unlike standard methods such as the widely used Maximum Likelihood Estimator (MLE), our methodology offers statistical guarantees, such as consistency and asymptotic normality, without relying on strong parametric assumptions. At the Gaussian model, our results (theoretical and obtained in simulations) show that the estimated parameters are similar to the ones obtained with the MLE. The computational performances of our approach have important practical implications. Indeed, the estimation of the parameters of large networks of thousands of GNSS stations (some of them being recorded over several decades) quickly becomes computationally prohibitive. Compared to standard methods, the processing time of the GMWMX is over $1000$ times faster considering time series recorded over $20$ years and allows the estimation of large scale problems within minutes on a standard computer. We validate the performances of our method via Monte-Carlo simulations by generating GNSS daily position time series with missing observations and we consider composite stochastic noise models including processes presenting long-range dependence such as power-law or Matérn processes. The advantages of our method are also illustrated using real time series from GNSS stations located in the Eastern part of the USA. 

\vspace{0.25cm}

\end{abstract}

\section{Introduction}
\label{sec:intro}

The Global Navigation Satellite System (GNSS) is an important tool to observe and model geodynamic processes such as post-glacial rebound (see e.g. ~\citealp{milne2001space}), hydrological loading (see e.g.~\citealp{bevis2002technical, tregoning2009atmospheric}) or crustal deformations (see e.g.~\citealp{williams2003offsets}). In the last three decades, the precision of the GNSS measurements was tremendously increased and allows researchers to study such geophysical signals with many details through a careful analysis of daily time series of GNSS receiver coordinates \citep{bock2016physical,Herring2016,he2017review}. Many geophysical applications focus on the estimation of the tectonic rate~\citep{Bock.1997,bock2016physical} either as a linear function \citep{fernandes2004angular,Bos2020}, or as a non-linear trend including offsets \citep{nielsen2013vertical,Blewitt2016}. To that end, the daily position time series are described as the sum of a noise and a geophysical signal. The latter can again be divided into station displacements due to geophysical phenomena (e.g., seasonal variations, tectonic movements, post-seismic relaxations of the crust) and other factors (e.g., small amplitude transient signals due to various disturbances \citealp{he2017review, rs13224523}).

\cite{bevis2014} are the first to suggest that the equations used to describe the motion of GNSS stations should be thought of as functional (or trajectory) models in crustal motion geodesy. This approach has also been applied to various fields such as gravity time series \citep{VanCamp2005}, mean sea level records \citep{Burgette2013, Montillet2020Springer}, and bridge oscillations \citep{OmidalizarandiHerrmannKargollMarxPaffenholzNeumann+2020+327+354}. In this contribution, we follow \cite{bevis2014} and \cite{he2019investigation}, and describe the geodetic time series by a functional and a stochastic noise model. We focus on obtaining the suitable parameter estimates together with reasonable associated uncertainties \citep{langbein2008noise,teferle2008continuous,bos2010influence,he2019investigation,Bevis2020,He2021}. The joint estimation of both deterministic and stochastic models is often based on the Maximum Likelihood Estimator (MLE) and has been implemented in various software packages such as CATS \citep{williams2008cats}, Est\_noise \citep{langbein2008noise} and Hector \citep{Bos2008}. Other methods use the Markov chain Monte-Carlo \citep{Olivaresandteferle2013} or the Expectation-Maximization (EM) algorithm \citep{Kargoll2020}. 

Unfortunately, the computational aspects related to the parameter estimation is often a key challenge when considering large datasets and/or complex stochastic noise models. Generally, various matrix operations are needed to compute the likelihood function which can become rapidly cumbersome for longer and longer time series. Powerful computing facilities (e.g., parallel processing, national computing centers) are required in order to process hundreds of stations, with some of them recording observations over several decades, in a reasonable amount of time. To speed up the processing time, several approximations of the MLE have been proposed. \cite{Bos2008, bos2013fast} reduced the computation time of a factor of $10$ to $100$ compared to the standard MLE method (depending on the length of the real time series) initially developed by \cite{williams2008cats}. \cite{Tehranchi2021} further improved the computational aspect of the method using restricted MLE. Despite these computational improvements, the analysis of crustal deformation or geodynamical activity on a large scale that (i) includes hundreds to thousands of GNSS stations \citep{He2021}, (ii) with some of them recording more than $25$ years of continuous observations and (iii) when different noise models must be tested, becomes impractical due to the large amount (e.g., weeks) of processing time required \citep{he2019investigation,Bos2020}. 

In this contribution, we propose a semi-parametric, computationally efficient and scalable method to estimate the parameters of linear models with dependent residuals. The key advancement of this new approach is that it avoids the use of strong parametric assumptions and drastically reduces the computational time required to estimate the models commonly used to describe GNSS time series data. Our method relies on a two-step statistical procedure which considers a (weighted) least squares approach to estimate the functional part of the model while the stochastic part of the model is obtained using the Generalized Method of Wavelet Moments (GMWM) proposed in \cite{guerrier2013wavelet}. We call our method the Generalized Method of Wavelet Moments with eXogenous inputs, or GMWMX. Interestingly, the Least Squares Variance Component Estimation (LS-VCE) proposed originally by \cite{teunissen2004towards}, and further developed and elaborated by \cite{Teunissen2008} and \cite{amiri2007least} is related to the proposed approach. Indeed, the LS-VCE is also a moment-based semi-parametric method with desirable computational properties. However, this method typically assumes that the covariance matrix of the observations is a linear combination with unknown variance components of known cofactor matrices. This requirement is relaxed in the GMWMX which allows to consider a large class of time series that cannot be expressed as linear combinations of known cofactor matrices, as shown in the following sections.

We test our method against the MLE using the Hector package developed by \cite{Bos2008} and \cite{he2019investigation}, a standard software to analyze geodetic time series. We focus especially on the processing time as a function of the length of the time series and the accuracy of the estimated geophysical parameters considering different stochastic noise models (e.g., a combination of power-law and white noise). Our analysis includes simulated and real GNSS daily position coordinate time series. The real data are provided by a few selected GNSS stations located in the east coast of the USA. We compare our estimates with (i) Hector's solutions and (ii) the velocity estimates provided by the Plate Boundary Observatory  (PBO - UNAVCO) \citep{Herring2016}.

This paper is organized as follows. The next section introduces the mathematical notations and a summary on the MLE. Section \ref{sec:proposed} derives the new estimator and discusses the contribution of the GMWMX with a specific application to GNSS daily position time series analysis. We then compare the results from our new estimator to the one obtained with the Hector software package for simulated and real observations in Sections \ref{sec:simu} and \ref{sec:study} respectively. We conclude with a discussion on the use of the GMWMX in environmental geodesy in Section~\ref{sec:conclusions}.

\section{Problem Formulation}
\subsection{Generalities and Notations} 
\label{sec:problem:formulation}

Throughout this paper we employ the following notations. For a vector $\mathbf{a} \in \real^n$ we define $\mathbf{a}_{i}$ as $i$-th element of the vector $\mathbf{a}$, for $i = 1, \ldots, n$. Similarly, for a matrix $\mathbf{A} \in \real^{n \times m}$, we define $\mathbf{A}_{i,j}$ as the $(i,j)$-th element of the matrix $\mathbf{A}$, for $i = 1, \ldots, n$ and $j = 1, \ldots, m$, and we denote $\mathbf{A}_i$ as the $i$-th row of $\mathbf{A}$. Given two matrices $\mathbf{A}, \, \mathbf{B} \in \real^{n \times m}$, we write $\mathbf{A} \, \propto \, \mathbf{B}$ to denote that $\mathbf{A}$ is proportional to $\mathbf{B}$ in the sense that there is a non-zero constant $k$ such that $\mathbf{A} = k \mathbf{B}$. Similarly, we write  $\mathbf{A} \, \cancel{\propto} \,\, \mathbf{B}$ to denote that $\mathbf{A}$ is not proportional to $\mathbf{B}$. Moreover, we write $\mathbf{A} > 0$ and $\mathbf{A} \geq 0$ to denote that the matrix $\mathbf{A}$ is positive definite and semi-positive definite, respectively. Finally, we use the notation $\overset{p}{\longrightarrow}$ and $\overset{d}{\longrightarrow}$ to denote convergence in probability and in distribution, respectively.

This work aims at developing a statistical inferential framework for the parameters of linear regression models with correlated residuals. While the method proposed in this article is generally applicable to various regression problems we consider in particular the models used for GNSS (daily) position time series. More precisely, we assume that the observations are generated from the following model
\begin{equation}
    \mathbf{Y} = \mathbf{A} {\x}_0 + \bm{\varepsilon},
    \label{eq:model}
\end{equation}
where $\mathbf{Y} \in \real^n$ denotes the response variable of interest (i.e., vector of GNSS observations), $\mathbf{A} \in \real^{n \times p}$ a fixed design matrix, ${\x}_0 \in \mathcal{X} \subset \real^p$ a vector of unknown constants and $\bm{\varepsilon} \in \real^n$ a vector of (zero mean) residuals. In many applications, ${\x}_0$ is of interest as it is related to, for example, the local tectonic rate and seismic phenomena (see e.g., \citealp{bock2016physical}). A common formulation of the functional component of the model is given by~\cite{he2017review}, which can be expressed for the $i$-th component of the vector $\mathbf{A} {\x}_0$ as follows:
\begin{equation}
    \mathbb{E}[\mathbf{Y}_i] = \mathbf{A}_i^T {\x}_0 = a+b\left(t_{i}-t_{0}\right)+\sum_{h=1}^{2}\left[c_{h} \sin \left(2 \pi f_{h} t_{i}\right)+d_{h} \cos \left(2 \pi f_{h} t_{i}\right)\right] + \sum_{k=1}^{n_{g}} g_{k} H\left(t_{i}-t_{k}\right),
    \label{eq:functionalmodel}
\end{equation}
where $a$ is the initial position at the reference epoch $t_0$, $b$ is the velocity parameter, $c_k$  and $d_k$ are the periodic motion parameters ($h = 1$ and $h = 2$ represent the annual and semi-annual seasonal terms, respectively). The offset terms models earthquakes, equipment changes or human intervention in which $g_k$ is the magnitude of the change at epochs $t_k$, $n_g$ is the total number of offsets, and $H$ is the Heaviside step function. Moreover, we assume that $\bm{\varepsilon}_i=\mathbf{Y}_i-\mathbb{E}[\mathbf{Y}_i]$ is a strictly (intrinsically) stationary process and that
\begin{equation}
    \bm{\varepsilon} \sim \mathcal{F} \left\{\mathbf{0}, \bm{\Sigma}(\bm{\gamma}_0)\right\} ,
    \label{eq:model:noise}
\end{equation}
where $\mathcal{F}$ denotes some probability distribution in $\real^n$ with mean $\0$ and covariance $\bm{\Sigma}(\bm{\gamma}_0)$. We assume that $\bm{\Sigma}(\bm{\gamma}_0) > 0$ and that it depends on the unknown parameter vector $\bm{\gamma}_0 \in \bm{\Gamma} \subset \real^q$. This parameter vector specifies the covariance of the observations and is often referred to as the \textit{stochastic parameters}. The formulation of the noise structure of $\bm{\varepsilon}$ is very general and includes a large class of time series models such as (the sum of) AutoRegressive Moving-Average (ARMA) models with additional noise, rounding errors and/or processes with long-range dependence. For example, this class of models includes the model considered in \cite{he2017review} by assuming $\mathcal{F}$ to be a multivariate normal distribution and that $\varepsilon_t = Z_t + R_t + U_t$, where $Z_t$ represents a Mat\'ern process (see e.g., \citealp{lilly2017fractional}), $R_t$ denotes a fractional (Gaussian) noise (see e.g. \citealp{li2006rigorous}) and $U_t$ represents a standard Gaussian white noise process. In practice, the estimation of $\bm{\gamma}_0$ is of interest as it could be informative regarding soil properties, such as moisture and groundwater depletion (see e.g., \citealp{bevis2005seasonal}), as well as atmospheric properties, which are of importance in climate change studies \citep{woppelmann2009rates}.

For simplicity, we let $\bm{\theta}_0 = \left[\bm{\x}_0\trans \;\; \bm{\gamma}_0\trans\right]\trans \in \bm{\Theta} = \mathcal{X} \times \bm{\Gamma} \subset \real^{p + k}$ denote the unknown parameter vector of the model described in~\eqref{eq:model}. The main goal of this paper is to propose a \textit{computationally efficient} inferential framework for $\bm{\theta}_0$ which enjoys desirable statistical properties while avoiding the specification of the probability distribution $\mathcal{F}$. Throughout, we consider a general class of probability distributions $\mathcal{F}$, which can be characterized by a set of mild regularity conditions specified later in Section~\ref{sec:proposed}. 

\subsection{Standard Likelihood Based Approach}
\label{sec:MLE}

The standard approach for the estimation of the problem defined in~\eqref{eq:model} is based on the MLE (see e.g.,~\citealp{Bos2008}) or closely related estimators such as the  Restricted MLE (see e.g.,~\citealp{Tehranchi2021}). In this section, we briefly review how maximum likelihood estimators can be constructed in this setting. Under the parametric assumption that the probability distribution $\mathcal{F}$ considered in~\eqref{eq:model:noise} is a multivariate normal distribution, the likelihood function for a generic $\bm{\theta} \in \bm{\Theta}$ is simply given by
\begin{equation}
    L\left(\bm{\theta} | \mathbf{Y} \right) = \exp\left\{-\frac{1}{2} \left(\mathbf{Y} - \mathbf{A}\x\right)\trans {\bm{\Sigma}\left(\bm{\gamma}\right)}^{-1}\left(\mathbf{Y}-\mathbf{A}\x\right)\right\} \left[(2\pi)^n \det\left\{\bm{\Sigma}\left(\bm{\gamma}\right)\right\}\right]^{-1/2},
    \label{eq:likelihood}
\end{equation}
allowing to define the MLE for $\bm{\theta}_0$ as
\begin{equation}
    \widehat{\bm{\theta}} = \left[\widehat{\x}\trans \;\; \widehat{\bm{\gamma}}\trans\right]\trans = \underset{\bm{\theta} \in \bm{\Theta}}{\argmax}\; L\left(\bm{\bt} | \mathbf{Y} \right) .
    \label{eq:mle}
\end{equation}
Under standard regularity conditions (see e.g. \citealp{newey1994large}), this estimator enjoys some desirable statistical properties such as consistency and asymptotic normality. In particular, under usual smoothness and mixing conditions it can be shown that
\begin{equation}
    \sqrt{n} \left( \widehat{\x} - \x_0 \right) \overset{{d}}{\longrightarrow} \mathcal{N}\left(\mathbf{0}, \mathbf{V}   \right), \;\;\; \text{where} \;\;\;  \mathbf{V} = \lim_{n \to \infty} n \left\{\mathbf{A}\trans  \bm{\Sigma}(\bm{\gamma}_0)^{-1} \mathbf{A} \right\}^{-1} .
    \label{eq:asym:mle}
\end{equation}
The estimator $\widehat{\x}$ is \textit{asymptotically optimal} since Aitken's Theorem (or more precisely its generalization given in \citealp{hansen2022modern}) shows that, in general, \textit{any} estimator of $\x$ (even in the case where $\bm{\Sigma}(\bm{\gamma}_0)$ is known), say $\bar{\x}$, is such that
\begin{equation}
    \label{eq:aitken}
    \var\left(\bar{\x}\right) \geq \left\{\mathbf{A}\trans  \bm{\Sigma}(\bm{\gamma}_0)^{-1} \mathbf{A} \right\}^{-1}.
\end{equation}
An important limitation of the MLE is the computational burden it often entail. Indeed, solving~\eqref{eq:mle} typically requires to evaluate the likelihood function in~\eqref{eq:likelihood} a large number of times. Each evaluation involves the inversion of the $n \times n$ matrix $\bm{\Sigma}(\bm{\gamma}_0)$ in~\eqref{eq:likelihood}, which is computationally expensive and can become problematic for large sample sizes. 

Alternatively, the Kalman filter can be used together with the EM algorithm to compute $\widehat{\bm{\theta}}$ while avoiding the matrix operations presented in~\eqref{eq:likelihood} (see \citealp{dempster1977maximum, shumway1982approach,shumway2000time}). While this approach can provide a viable solution in some cases, the ``M'' step can be very complex while the ``E'' step is often computationally cumbersome, therefore finding the MLE is not always a simple task. Moreover, this approach becomes particularly challenging when $n$ is large and/or when the model describing $\varepsilon_i$ is complex such as a sum of latent random processes as presented, for example, in Section~\ref{sec:problem:formulation}. The limited practical applicability of the MLE in this context was, for example, illustrated in \cite{stebler2014wavelet}.

Furthermore, the MLE presented in this section and considered, for example, in \cite{Bos2020} is based on the strong parametric assumption (often referred to as the Gauss–Markov hypothesis) that the noise $\bm{\varepsilon}$ follows a multivariate Gaussian distribution. This hypothesis guarantees the reliability of the estimated functional and stochastic models via MLE but it also implies the following:

\begin{enumerate}
    \item The mean of the correlated noise varies slowly with time. We then rule out the occurrence of specific events of a large amplitude, such as aggregations or bursts of spikes (i.e. intermittency), which could invalidate such an assumption.
    \item We exclude any potential misspecification in the functional model selection. For example, missing geophysical information (e.g., offsets) can generate heavy-tailed and/or skewed distributions of the residuals of GNSS time series models \citep{npg-28-121-2021}. 
\end{enumerate}

Consequently, the MLE presented in this section relies on strong parametric assumptions which are often difficult to verify in practice and often appear unrealistic due to the presence of large mean deviations in GNSS time series.

\section{The Generalized Method of Wavelet Moments with Exogenous Inputs}
\label{sec:proposed}

\begin{figure}[ht]
    \centering
    \includegraphics{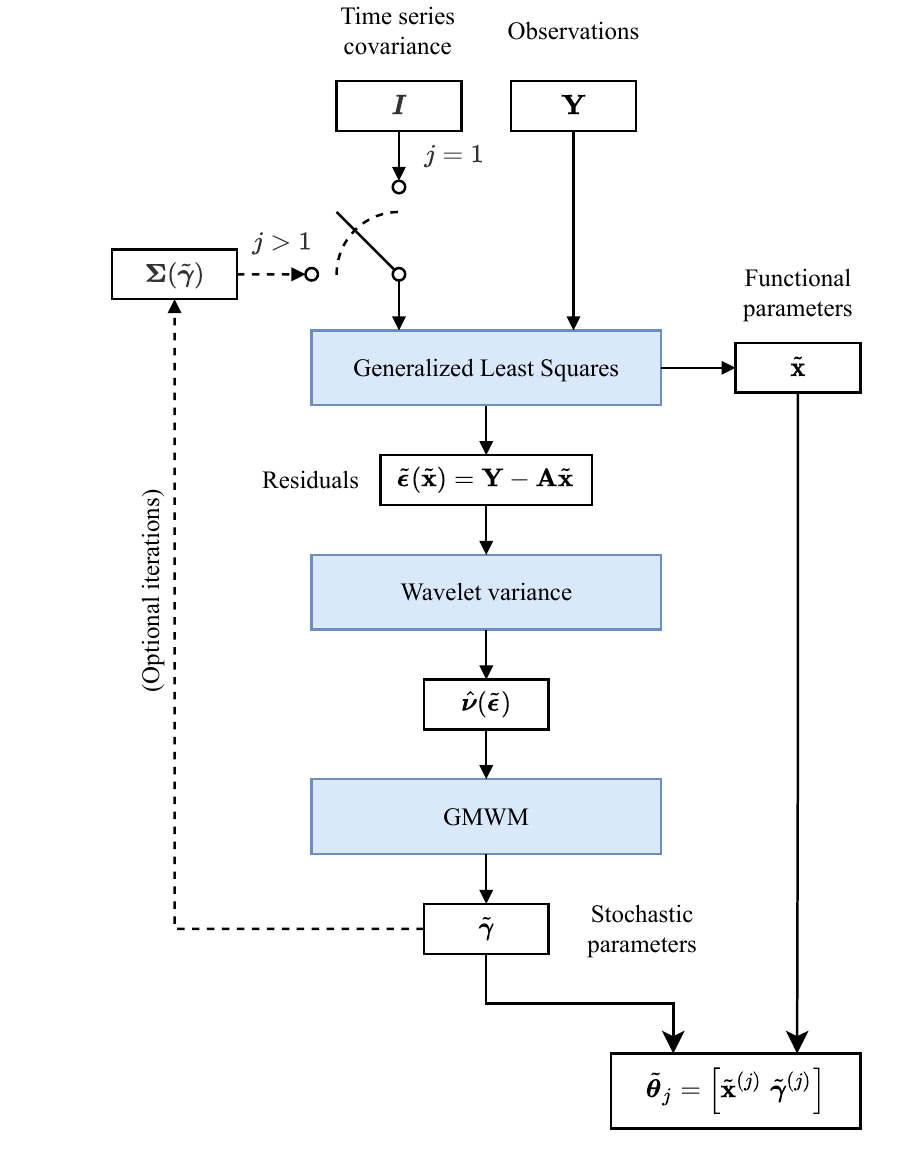}
    \caption{Flowchart representing the GMWMX method described in Section~\ref{sec:stat:part}.}
    \label{fig:flowchart}
\end{figure}

In this section we introduce the GMWMX approach which extends the standard GMWM of \cite{guerrier2013wavelet} in the context of linear regression with correlated residuals. This method can be applied, for example, to the estimation of the parameters of the model described in~\eqref{eq:model}. The proposed approach is computationally efficient and allows to considerably alleviate the computational limitations of standard methods such as the MLE. Unlike methods relying on fully specified parametric model, we relax some of the requirements imposed on $\mathcal{F}$. Indeed, we only require that $\bm{\varepsilon}_t$ is a strictly (intrinsically) stationary process with finite fourth moment and covariance matrix $\bm{\Sigma}(\bm{\gamma}_0)$. Thus, the GMWMX is a \textit{semi-parametric} method in the sense that its statistical properties are preserved for a general class of probability distributions $\mathcal{F}$ (which can remain unspecified). Compared to fully parametric methods such as the MLE, our approach provides statistical guarantees \textit{for all} zero-mean probability distributions (with finite fourth moment) and covariance matrix $\bm{\Sigma}(\bm{\gamma}_0)$. Moreover, the GMWMX is a semi-parametric approach based on the principle of Generalized Least Squares (GLS) combined with the GMWM framework. Indeed, our approach considers initially a coarse approximation of $\bm{\Sigma} (\bm{\gamma}_0)$ as defined in~\eqref{eq:model:noise}, which is used in a GLS approach to obtain an estimate of $\x_0$. From this estimate, we then compute a GMWM-based estimator of $\bm{\gamma}_0$. Our framework allows to iterate this process in order to improve statistical efficiency. The procedure is schematically depicted in Figure~\ref{fig:flowchart}, formally defined in Section~\ref{sec:stat:part} and its benefits are summarized in Section~\ref{sec:main:results}.

\subsection{Proposed Statistical Framework}
\label{sec:stat:part}

The GLS is a common method used to estimate the unknown parameters in a linear regression model with correlated residuals assuming that the covariance matrix $\bm{\Sigma}(\bm{\gamma}_0)$ of $\bm{\varepsilon}$ is known. In our setting, this requirement is not realistic but we let $\bm{\Sigma}$ denote the \textit{assumed} covariance matrix of $\bm{\varepsilon}$. The notation $\bm{\Sigma}$ is used to highlight that $\bg_0$ (and thus $\bm{\Sigma}(\bg_0)$) is unknown allowing to consider different approximations of $\bm{\Sigma}(\bg_0)$ by $\bm{\Sigma}$. Based on the assumed covariance $\bm{\Sigma}$ we obtain the following GLS estimator:
\begin{equation}
\widetilde{\x}\left(\bm{\Sigma}\right) =  \, \underset{\x \in \mathcal{X}}{\argmin}\; \left\{\Y - \A\x\right\}^T \bm{\Sigma}^{-1} \left\{ \Y - \A\x\right\}
= \left(\A\trans \bm{\Sigma}^{-1} \A\right)^{-1} \A\trans  \bm{\Sigma}^{-1} \Y.  
\label{eq:gls:x}
\end{equation}
In case where we consider the crude approximation $\bm{\Sigma}\, \propto \, \I$, the estimator reduces to the
ordinary least squares estimator and we obtain: 
\begin{equation}
    \widetilde{\x} = \widetilde{\x}\left( \I\right) =  \left( \A\trans \A\right)^{-1} \A\trans \Y.
    \label{eq:ls}
\end{equation}
This estimator is simple to compute and enjoys well-known statistical properties (as discussed later in this section). Indeed, under very mild conditions based on functional dependence measures as proposed initially by \cite{wu2005nonlinear}, we have that $\widetilde{\x}$ is consistent for $\x_0$ and admits the following limiting distribution (see Theorem~1 of \citealp{wu2007}):
\begin{equation}
    \sqrt{n} \left( \widetilde{\x} - \x_0 \right) \overset{d}{\longrightarrow} \mathcal{N}\left(\mathbf{0}, \mathbf{V}^{\ast}   \right), \;\;\; \text{where} \;\;\;  \mathbf{V}^{\ast} = \lim_{n \to \infty} n \left(\mathbf{A}\trans  \mathbf{A} \right)^{-1}  \mathbf{A}\trans   \bm{\Sigma}(\bm{\gamma}_0) \mathbf{A} \left(\mathbf{A}\trans  \mathbf{A} \right)^{-1}.
\label{eq:step1:limit}
\end{equation}
However, the estimator is not asymptotically optimal and compared to the MLE $\widehat{\bm{\x}}$, for $\bm{\Sigma}(\bm{\gamma}_0)\, \cancel{\propto}\,\, \mathbf{I}$ we have:
\begin{equation}
   \lim_{n \to \infty} \;   \var\left\{\sqrt{n} \left( \widetilde{\bm{\x}} - \bm{\x}_0 \right)\right\} - \var\left\{\sqrt{n} \left( \widehat{\bm{\x}} - \bm{\x}_0 \right)\right\} = \mathbf{V}^{\ast} - \mathbf{V} > 0.
   \label{eq:pos:def:mat}
\end{equation}
The derivation of this result is given in Appendix~\ref{app:derivation:1}. This result implies that any linear combination of $\wt\x$ has larger asymptotic variance with respect to the same linear combination of $\wh\x$. Therefore, $\widetilde{\bm{\x}}$ is asymptotically less efficient than the MLE $\wh{\bm{\x}}$ in the case of correlated and/or heteroscedastic residuals (i.e., $\bm{\Sigma}(\bm{\gamma}_0) \, \cancel{\propto}\,\, \mathbf{I}$).

Based on a suitable estimator of $\x_0$, such as $\widetilde{\bm{\x}}$, we can compute the (estimated) residuals of model~\eqref{eq:model} whose population-level version of $\bm{\varepsilon}$ is defined as
\begin{equation}
    {\bm{\varepsilon}}\left(\x\right) = \mathbf{Y} - \A \x,
    \label{eq:residuals}
\end{equation}
and a natural estimator of $\bm{\varepsilon}$ is simply ${\wt{\bm{\varepsilon}}} = \bm{\varepsilon}\left(\wt{\x}\right)$. This estimator is consistent for $\bm{\varepsilon}(\x_0)$ since $\wt\x$ is consistent for $\x_0$ as implied by~\eqref{eq:step1:limit} and the continuous mapping theorem. More precisely, we have for all $i \in \{1, \ldots, n\}$
\begin{equation*}
   {\wt{\bm{\varepsilon}}}_i = \bm{\varepsilon}_i\left(\wt{\x}\right)  = \Y_i - \mathbf{A}_i\trans \wt{\x}  \overset{p}{\longrightarrow}  \Y_i - \mathbf{A}_i\trans {\x_0} = \bm{\varepsilon}_i(\x_0) = \bm{\varepsilon}_i.
\end{equation*}
The vector of residuals ${\wt{\bm{\varepsilon}}}$ allows to construct an estimator of $\bg_0$ using the GMWM methodology. The latter is an estimation framework that allows to consider a wide range of models including some complex (latent) models where standard methods typically fail due to the model complexity and/or the unrealistic computational burden they entail (see e.g. \citealp{stebler2011constrained, stebler2014wavelet}). In short, this approach uses a quantity called Wavelet Variance (WV) (see e.g., \citealp{percival2000wavelet}) in the spirit of a Generalized Method of Moments (GMM) estimator of \cite{hansen1982large}. The GMWM estimator based on an estimator of $\x_0$, say $\x$, is defined as follows:
\begin{equation}
    \wt{\bm{\gamma}}\left(\x\right) = \underset{\bm{\gamma} \in \bm{\Gamma}}{\argmin}\; \left\{\wh{\bm{\nu}}\left({\x}\right) - \bm{\nu}(\bm{\gamma})\right\}\trans\bm{\Omega} \left\{\wh{\bm{\nu}}\left({\x}\right) - \bm{\nu}(\bm{\gamma})\right\},
    \label{eq:step2}
\end{equation}
where $\bm{\nu}(\bm{\gamma})$ is the WV vector implied by the model. This quantity is an explicit function of the parameters for a large class of models based on the general results of \cite{zhang2008allan}. The vector $\wh{\bm{\nu}}\left({\x}\right)$ denotes the estimated Haar WV computed on ${\bm{\varepsilon}}\left({\x}\right)$ and $\bm{\Omega}$ corresponds to an appropriate (possibly estimated) positive-definite weighting matrix (see e.g., \citealp{guerrier2013wavelet} for more details). Additional details on these quantities are given in Appendix~\ref{app:gmwm}. Using the previously defined quantities, the idea behind the GMWM estimator presented in~\eqref{eq:step2} is to match $\wh{\bm{\nu}}\left({\x}\right)$ with $\bm{\nu}(\bm{\gamma})$ in a GLS fashion. This estimator is consistent and asymptotically normally distributed under arguably weak conditions (see \citealp{guerrier2013wavelet, guerrier2021robust} for details). By the continuous mapping theorem and for any consistent estimator of $\x_0$, say $\x$, we have under technical requirements (see \citealp{guerrier2021robust} for details) that $\wh{\bm{\nu}}\left(\x\right)$ is a consistent estimator of $\bm{\nu}(\bm{\gamma}_0)$. In particular, we propose to consider $\wt{\x}$ as defined in~\eqref{eq:ls} which satisfies
\begin{equation}
   \wh{\bm{\nu}}\left(\wt{\x}\right) \overset{p}{\longrightarrow}  \bm{\nu}(\bm{\gamma}_0),
\end{equation}
and, therefore, under the conditions of \cite{guerrier2013wavelet} we have
\begin{equation}
    \wt{\bm{\gamma}} = \wt{\bg}\left(\wt{\x}\right) \overset{p}{\longrightarrow}  \bm{\gamma}_0.
    \label{eq:limit:step2}
\end{equation}
Similarly to $\wt{\x}$, the estimator $\wt{\bg}$ is (asymptotically) less efficient than $\wh{\bg}$ as defined in~\eqref{eq:mle}. To narrow this gap, it is possible to consider instead the following procedure which iteratively recomputes $\wt{\x}$ defined in~\eqref{eq:gls:x} based on updated estimator of $\bm{\Sigma}(\bg_0)$. Starting at $j = 1$ with $\bm{\Sigma}^{(0)} = \mathbf{I}$, we define
\begin{equation}
    \begin{aligned}
    \label{eq:gmwm:reg}
    \wt{\x}^{(j)} &=
     \left\{\A\trans \left(\bm{\Sigma}^{(j-1)}\right)^{-1} \A\right\}^{-1} \A\trans  \left(\bm{\Sigma}^{(j-1)}\right)^{-1} \Y,\\
    \wt{\bm{\gamma}}^{(j)} &= \underset{\bm{\gamma} \in \bm{\Gamma}}{\argmin}\; \left\{\wh{\bm{\nu}}\left(\wt{\x}^{(j)}\right) - \bm{\nu}(\bm{\gamma})\right\}\trans\bm{\Omega} \left\{\wh{\bm{\nu}}\left(\wt{\x}^{(j)}\right) - \bm{\nu}(\bm{\gamma})\right\},\\
    \bm{\Sigma}^{(j)} &= \bm{\Sigma}\left(\wt{\bm{\gamma}}^{(j)}\right) = \var\left(\mathbf{Y} | \wt{\bm{\gamma}}^{(j)}\right).
    \end{aligned}
\end{equation}
Please see again Figure~\ref{fig:flowchart} for a schematic depiction of this iterative procedure.
In fact, we have that $\wt{\x}^{(j)}$ is asymptotically optimal for all $j \geq 2$ in the sense that
\begin{equation}
   \lim_{n \to \infty} \;  \var\left\{\sqrt{n} \left( \widehat{\x} - \x_0 \right)\right\} - \var\left\{\sqrt{n} \left( \wt{\x}^{(j)} - \x_0 \right)\right\} = \0.
   \label{eq:asym:gmwmx2}
\end{equation}
This result is a direct consequence of the consistency of $\wt{\bm{\gamma}}^{(j)}$ for $j \geq 1$, the continuous mapping theorem and Slutsky's theorem provided that the function $\bm{\Sigma}(\bg)$ is continuous in $\bg$. This is a plausible requirement which is satisfied for the majority of time series models. Equivalently to~\eqref{eq:asym:gmwmx2} we can write
\begin{equation}
    \sqrt{n} \left( \wt{\x}^{(j)} - \x_0 \right) \overset{{d}}{\longrightarrow} \mathcal{N}\left(\mathbf{0}, \mathbf{V}   \right),
\end{equation}
for $j \geq 2$ and where $\mathbf{V}$ is defined in~\eqref{eq:asym:mle}.

The procedure described in~\eqref{eq:gmwm:reg} is known as the \textit{iterated} GMM, when iterated until convergence. The special case of $j = 2$ is the so-called two-step GMM widely used in econometrics~\citep{greene2003econometric}. In this article our main focus is on providing a reliable yet computationally efficient estimator of $\bm{\theta}_0$. For this reason, we opt for the convenient choices of $j \in \{1, \,2\}$ which corresponds to the following estimators:
\begin{equation}
    \begin{aligned}
    \wt{\bt}_j &= \left[\wt{\x}^{(j) \rm T} \;\; \wt{\bm{\gamma}}^{(j) \rm T} \right]\trans.
    \end{aligned}
    \label{eq:def:gmwmx}
\end{equation}
These particular choices are consistent with the statistical properties of this estimator, but they are based mainly on our empirical experience and desire for simplicity and is not necessarily an optimal choice. The first estimator $\wt{\bt}_1$ is particularly computationally efficient and its computational complexity is only $\mathcal{O}\{\log_2(n) n\}$. The second $\wt{\bt}_2$ is slightly more computationally demanding but asymptotically optimal for $\wt{\x}^{(2)}$. Hereafter, we denote the estimator defined in~\eqref{eq:def:gmwmx}  with one or two iterations as the GMWMX-1 and the GMWMX-2, respectively. The performances of the proposed estimators are illustrated and discussed in details in Section~\ref{sec:simu}. As previously mentioned, the iterative procedure described in~\eqref{eq:gmwm:reg}, with $j > 2$, could be used to further improve the statistical properties of $\wt{\bt}_2$, but we do not pursue this direction here.

\subsection{Contributions}
\label{sec:main:results}

The general statistical framework proposed in the previous section has several advantages over the standard MLE. First, our approach is \textit{semi-parametric} in the sense that the probability distribution $\mathcal{F}$ considered in~\eqref{eq:model:noise} is left unspecified. Throughout, we consider a general class for the probability distribution $\mathcal{F}$, which can be characterized by a set of mild regularity conditions. These advantageous features avoid the common assumption that the residuals $\bm{\varepsilon}$ are issued from a multivariate normal distribution. Indeed, this assumption is often unrealistic in practice as the (estimated) residuals may have asymmetric and leptokurtic distributions. Consequently, our methodology offers statistical guarantees, such as consistency or asymptotic normality without relying on strong parametric assumptions.

Secondly, the proposed approach is \textit{computationally efficient} while preserving adequate statistical properties. The computational cost of our method is comparable to a single evaluation of the standard Gaussian likelihood function with its computational bottleneck corresponding to the inversion of an $n \times n$ matrix. Indeed, our two-step estimator $\wt{\bt}_2 = \left[\wt{\x}^{(2)\rm T} \;\; \wt{\bm{\gamma}}^{(2) \rm T} \right]$ defined in~\eqref{eq:def:gmwmx} is consistent for $\bm{\theta}_0$. Moreover, the estimator $\wt{\x}^{(2)}$ for $\x_0$ is \textit{asymptotically optimal} and corresponds to the (asymptotically) \textit{best unbiased estimator} in the sense of \cite{hansen2022modern}. The estimator $\wh{\bm{\gamma}}$ for $\bm{\gamma}_0$ has similar statistical properties to ones of the MLE (at the Gaussian model) but possibly comes at the price of a marginally inflated variance due to the semi-parametric nature of the procedure.

Moreover, our methodology is \textit{scalable} as it provides a simple strategy using $\wt{\bt}_1$ defined in ~\eqref{eq:def:gmwmx} to marginally reduce the statistical properties of our estimator in order to considerably limit the computational burden. Indeed, in situations where large networks of GNSS stations are considered, the computational cost can be further reduced to become comparable to the computation of the standard least squares estimator. Consequently, large scale problems can be solved within a few minutes on a standard computer. 

\section{Results and Discussions} 
\label{results}
\subsection{Simulation Studies}
\label{sec:simu}

In this section we evaluate via Monte-Carlo simulations the performances of the GMWMX-1 and GMWMX-2 estimators defined in~\eqref{eq:def:gmwmx} as well as the validity of their associated confidence intervals compared to the MLE as implemented in the software Hector v1.9 of \cite{Bos2008}. We consider several simulated scenarios based on~\eqref{eq:functionalmodel} for the functional model and different stochastic models. The impact of missing values is also investigated and each simulation is replicated to consider this case. For the stochastic model, we consider $\bm{\varepsilon}$ to be the sum of a power-law and a Gaussian white noise, which is a widely accepted model~\citep{zhang1997southern, Bos2008, Klos2014}. Another choice of the stochastic model is considered in Appendix~\ref{app:matern} where the power-law process is replaced with a Mat\'ern process. The values of the functional parameters are fixed as follows: $a = 0$, $b = 5$ mm/year, and the annual periodic motion has an amplitude of $2.5$ mm, and a phase with respect to the reference epoch of $145$ days. For the stochastic part, we consider $\sigma^2_\text{PL} = 10$ mm$^2$ and $d = 0.4$ while the variance of the white noise is $\sigma^2_\text{W} = 15$ mm$^2$/years$^2$. All our simulations are based on $N = 10^3$ Monte-Carlo replications and rely on the Hector package for data generation.

\begin{figure}[ht]
	\centering
	\includegraphics{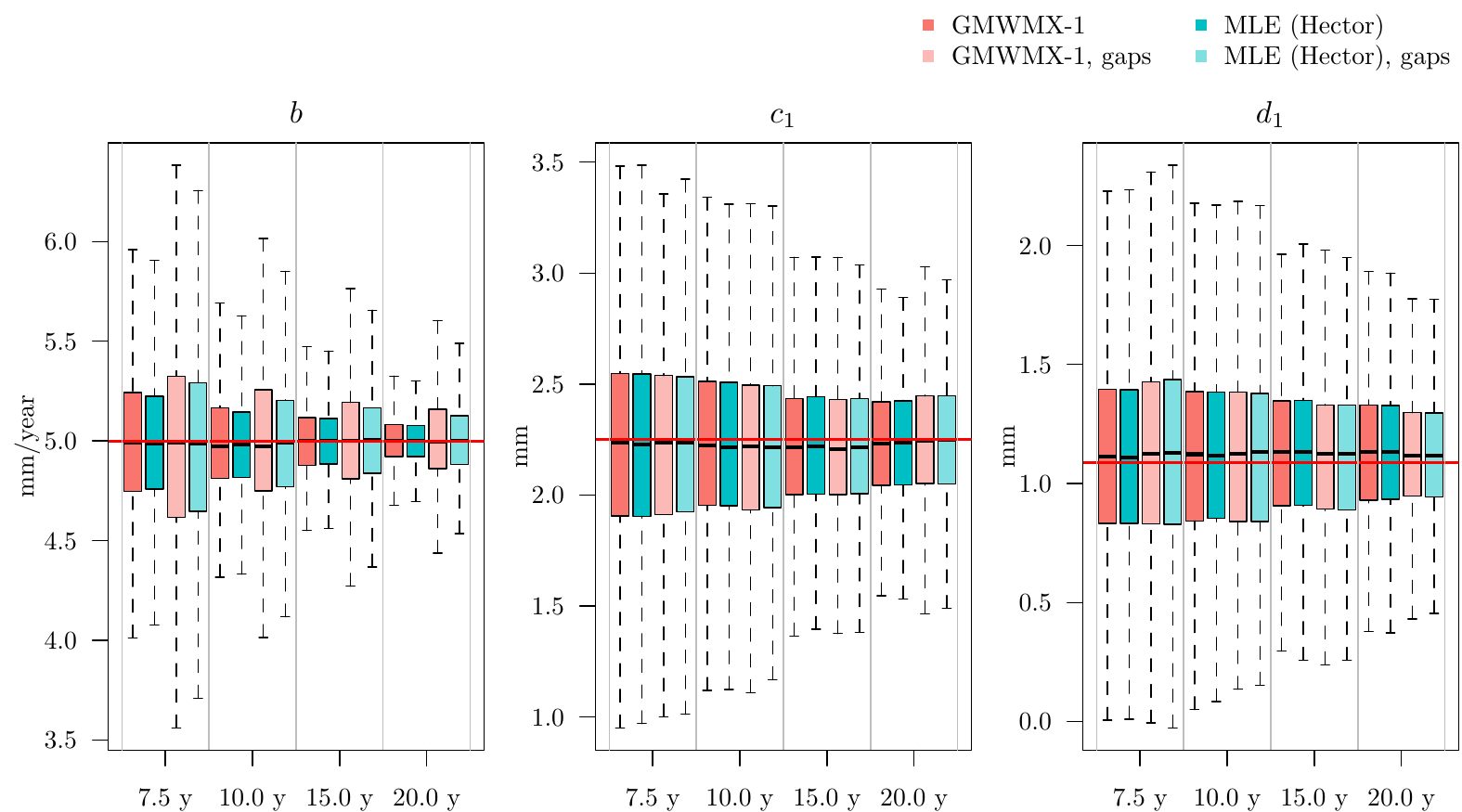}
	\caption{Boxplots of the empirical distribution of the estimated parameters obtained with GMWMX-1 and the MLE based on $10^3$ Monte-Carlo replications in the nominal scenario and the one including missing data and offsets. The parameters $b$, $c_1$ and $d_1$ correspond to the trend and the seasonal variation parameters as described in~\eqref{eq:functionalmodel}, respectively. For each parameter, the red line highlights the true value used in the data simulation.}
	\label{fig:detparams}
\end{figure}

\begin{figure}[ht]
	\centering
    \includegraphics{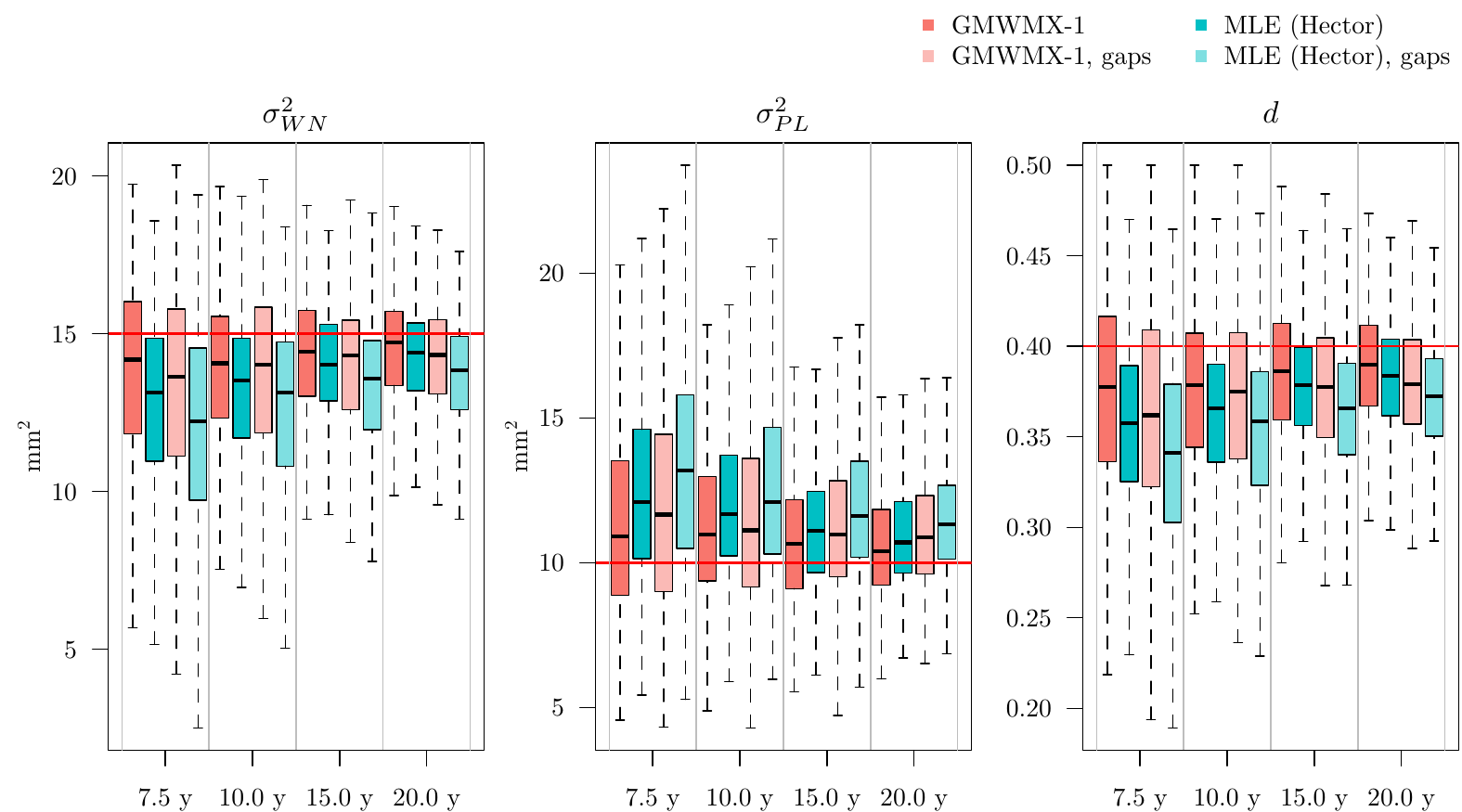}
	\caption{
	Boxplots of the empirical distribution based on $10^3$ Monte-Carlo replications of the estimated parameters obtained with GMWMX-1 and the MLE in the nominal scenario and the one including missing data and offsets. The parameters $\sigma^2_{WN}$ and $\sigma^2_{PL}$ are the white noise and the power-law variance, respectively, while $d$ is the power-law exponent. For each parameter, the red line highlights the true value used in the data simulation.}
	\label{fig:stochparams}
\end{figure}

In our first simulation setting, we compare the GMWMX-1 and the MLE with various sizes of daily observations, i.e., $7.5$, $10$, $15$ and $20$ years. For each length of GNSS coordinate time series, we consider a) the nominal scenario, where all data points are available and no offsets are present in the real time series ($n_g = 0$); b) a more realistic scenario in which 1) randomly selected $5$\% of the observations are missing, and 2) one random offset every $5$ years is introduced in the time series at known (but varying between Monte-Carlo replicates) epochs ($n_g = 2,2,3$ or $4$, depending on the available data), following~\cite{he2017review}. We indicate the results of the estimators compared in this second scenario by ``gaps''.  The amplitude of each offset, $g_k$, is sampled from a Gaussian distribution with zero mean and standard deviation of $10$~mm. The estimated parameters from the (geophysical) functional model and for the stochastic one are shown in Figures~\ref{fig:detparams} and \ref{fig:stochparams}, respectively. It can be observed how the functional parameters are estimated well by the two methods, albeit the GMWMX-1 exhibiting a slightly increased variance for the trend parameter $b$. Regarding the stochastic parameters, both methods tend to estimate a higher variance for the power-law noise at the expenses of the white noise. While still being small, this bias appears slightly larger for Hector, and in all cases decreases with the sample size. Similar results are presented in Appendix~\ref{app:matern}, where we consider a different stochastic model made of a combination of a Gaussian white noise and a Mat\'ern process~\citep{lilly2017fractional}.

\begin{figure}[ht]
	\centering
	\includegraphics{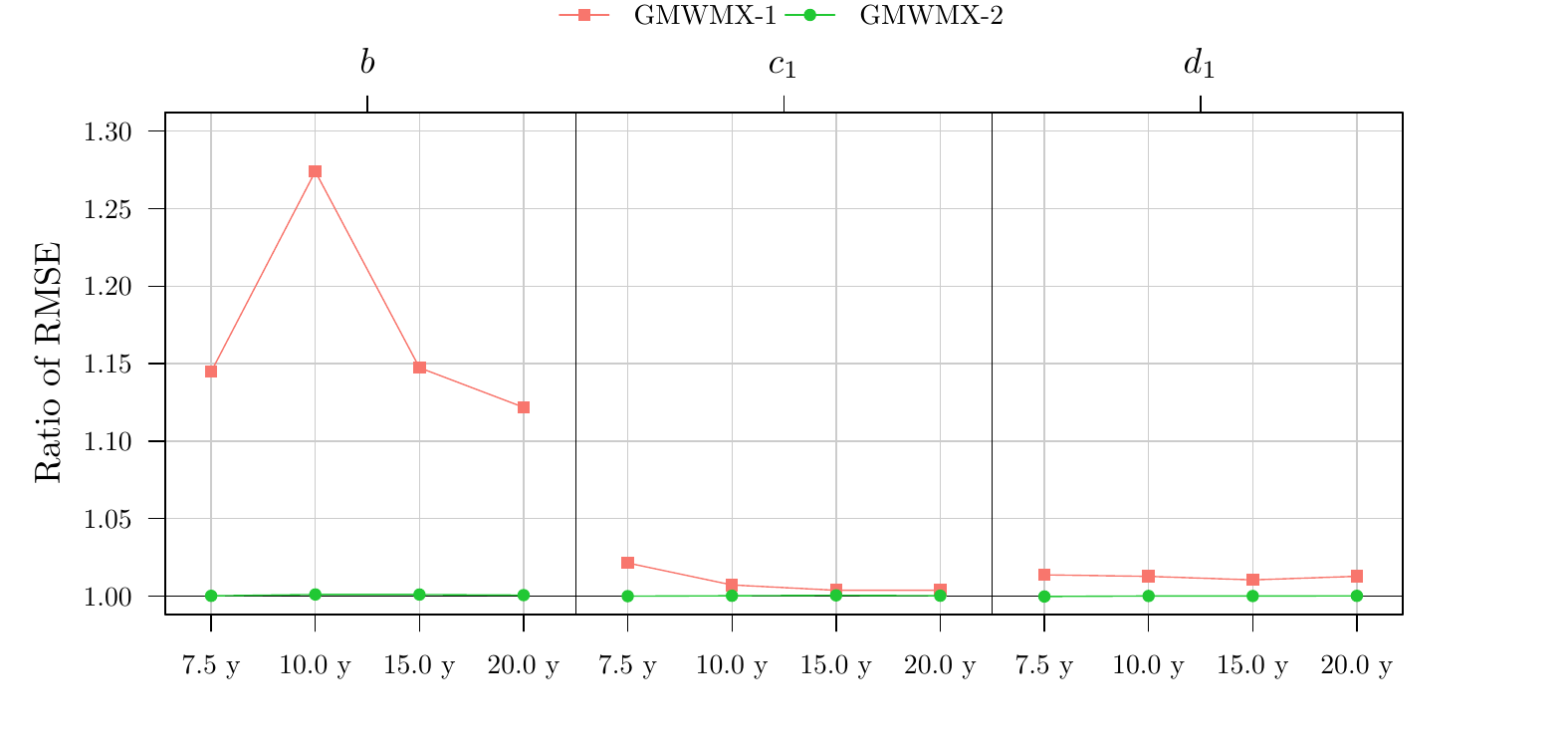} 
	\caption{Ratio of the estimated RMSE of the GMWMX-1 and GMWM-2 compared to the MLE for the functional parameters $b$, $c_1$ and $d_1$ as a function of the sample size.}
	\label{fig:rmse}
\end{figure}

Next, we consider the GMWMX-2 which is a more statistically efficient estimator than GMWMX-1. This second estimator is expected to have very similar finite sample performances to the MLE for the functional parameters due to their asymptotic equivalence presented in~\eqref{eq:asym:gmwmx2}. We focus on the previous nominal scenario and consider the ratio of the Root Mean Square Error (RMSE) of GMWMX-1 and GMWMX-2 over the MLE. The results are presented in Figure~\ref{fig:rmse} and are in line with the theory presented in Section~\ref{sec:stat:part}. Indeed, the GMWMX-2 appears to have an almost identical RMSE with respect to the MLE (the ratio being close to $1$) and it clearly improves over GMWMX-1. The improvement is more pronounced for the trend $b$ (up to $25$\%) and less evident for the seasonal parameters $c_1$ and $d_1$ (i.e. less than $5$\%).

\begin{figure}[ht]
	\centering
	\includegraphics[]{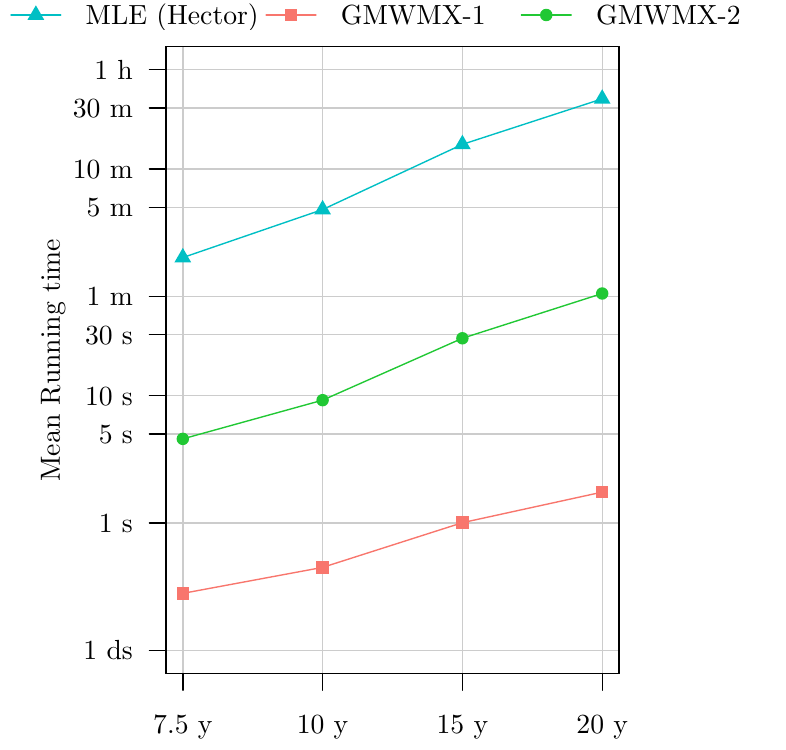} 
    \caption{Mean running time of the MLE, the GMWMX-1 and the GMWMX-2 as a function of the sample size.}
    \label{fig:runningtime}
\end{figure}

An important advantage of the proposed method is its computational efficiency with respect to the MLE, and thus a radically short running time. In Figure~\ref{fig:runningtime} we compare the running time of the \hbox{GMWMX-1} and the \hbox{GMWMX-2} with respect to the MLE as a function of the sample size. While Hector takes on average $2$ minutes for the smallest sample size of $7.5$ years, corresponding to $2,737$ data points, the estimation with GMWMX-1 takes on average less than $2$ seconds for the largest considered sample size of $20$ years. Therefore, the GMWMX-1 is between $400$ and $1,200$ times faster than the MLE in the cases considered in this simulation. Regarding the GMWMX-2, the increased statistical efficiency (i.e. lower asymptotic variance) comes at the price of a longer running time because of the need to compute the inverse of ${\bm{\Sigma}}(\tilde{\bg}^{(1)})$ once. However, the GMWMX-2 is still between $20$ to $40$ times faster than the MLE while providing statistically equivalent results as shown in Figure~\ref{fig:rmse}.

\begin{figure}[ht]
	\centering
	\includegraphics[]{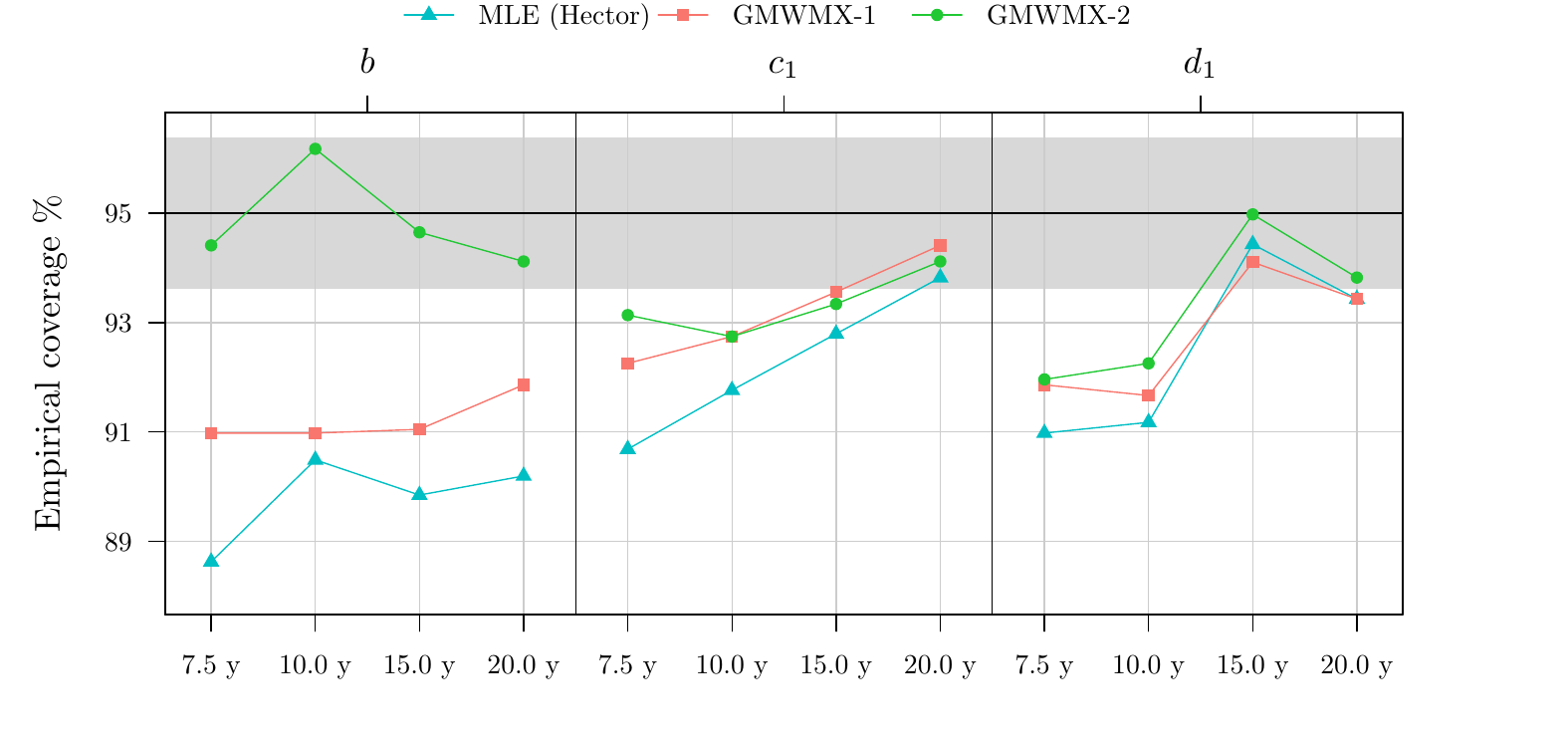} 
    \caption{Empirical coverage of the confidence intervals at level $1 - \alpha = 0.95$ for the functional parameters for GMWMX-1, GMWMX-2 and the MLE. The grey area represents a 95\% confidence intervals of the simulation errors based on $10^3$ Monte-Carlo replicates.}
    \label{fig:coverage}
\end{figure}

Another significant result is related to the validity of the confidence intervals that can be constructed for the functional parameters. The estimated uncertainty for each parameter allows to construct confidence intervals at any chosen confidence level, i.e., the intervals within which the true parameter should lie with the given probability. As a rule of thumb, if $\tilde \sigma$ is the estimated uncertainty for a given parameter which is asymptotically normally distributed, then the interval constructed around the estimated value $\pm 1.96 \tilde \sigma$ yields the approximately $95$\% confidence interval for that parameter (see Appendix~\ref{ap:ci} for details). With Monte-Carlo simulations, the true parameter values are known: this makes it possible to verify the validity of the constructed confidence intervals, i.e., if they include the true parameter value with the required probability. The empirical coverage of the deterministic parameters, defined as the proportion of simulations in which the true value of the parameters is inside the computed confidence intervals, is shown for the MLE, the GMWMX-1 and the GMWMX-2 in Figure~\ref{fig:coverage}. We observe that all methods yield empirical coverages close to the chosen confidence level of $95$\%. Therefore, the uncertainty for the functional parameters is correctly estimated by all methods and no significant bias is present. However, the GMWMX-2 appears to present empirical coverages that are closer to the chosen confidence level of $95$\%. In particular, the GMWMX-2 provides more accurate confidence intervals for the trend parameter $b$ for all sample sizes with respect to both the GMWMX-1 and the MLE in the considered case. These results  may be explained by the smaller bias of the GMWMX-2 for the stochastic parameters (as shown in Figure~\ref{fig:stochparams}) compared to the MLE. 

Similar conclusions can be obtained with other stochastic models for the residuals $\bm{\varepsilon}$, such as a white noise summed with a Mat\'ern process shown in Appendix~\ref{app:matern}. Further simulation studies suggest that the GMWMX-2 yields confidence intervals with marginally better empirical coverage with respect to the MLE or the GMWMX-1 when the residuals $\bm{\varepsilon}$ do not follow a multivariate Gaussian distribution (e.g. skewed Student's t-distribution). However, the inferential advantages of the proposed semi-parametric method outside of the Gaussian model is beyond the scope of our study and is left for further research.

\subsection{Case Study}
\label{sec:study}

\begin{figure}
	\centering
	\includegraphics[]{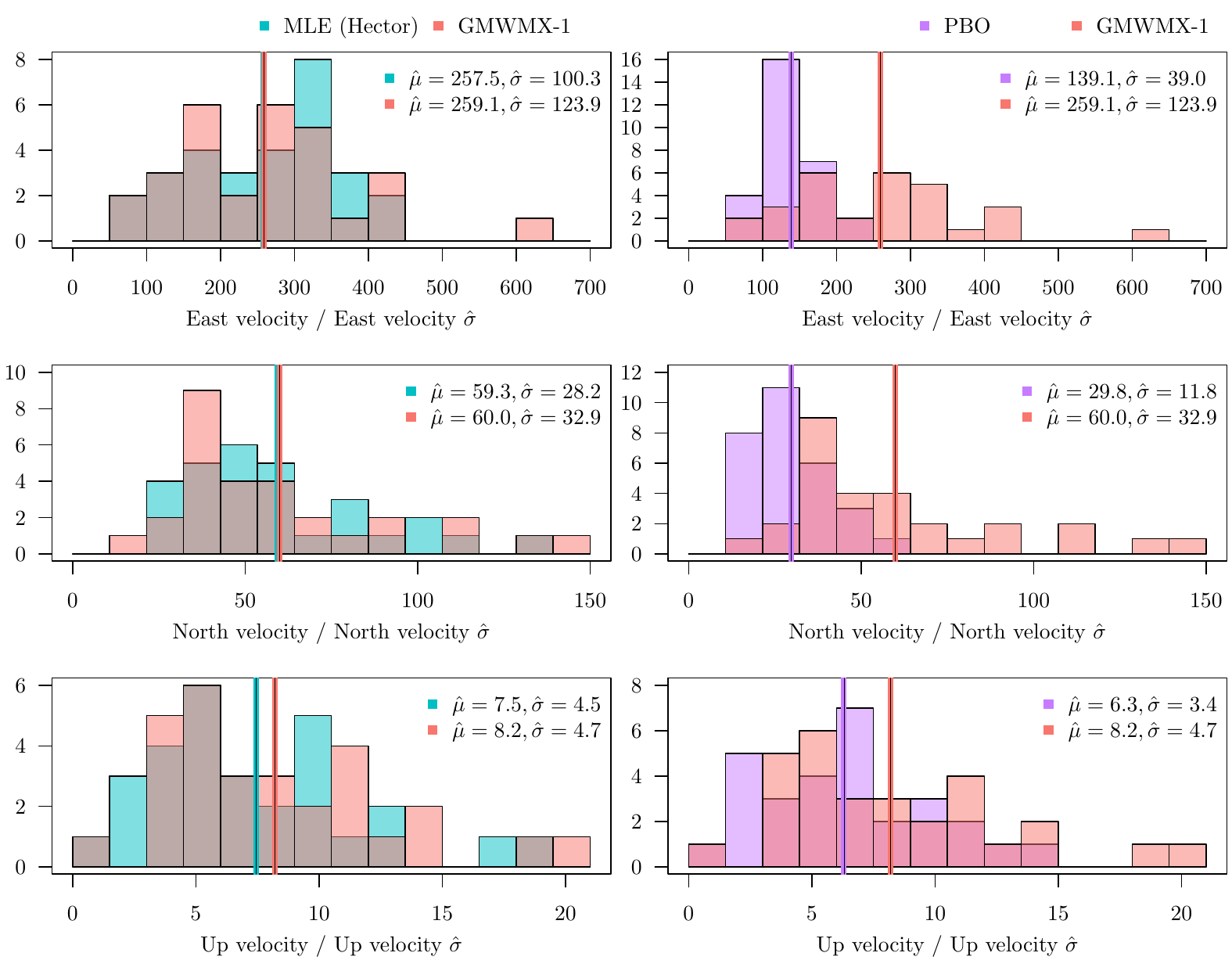}
	\caption{Ratios between estimated North-East velocities and crustal uplift divided by their respective estimated standard deviation for the GMWMX-1, the MLE and the PBO product for $33$ GNSS receivers distributed over the East coast of the USA.}
	\label{fig:ratios}
\end{figure}

\begin{figure}[ht]
	\centering
	\includegraphics[]{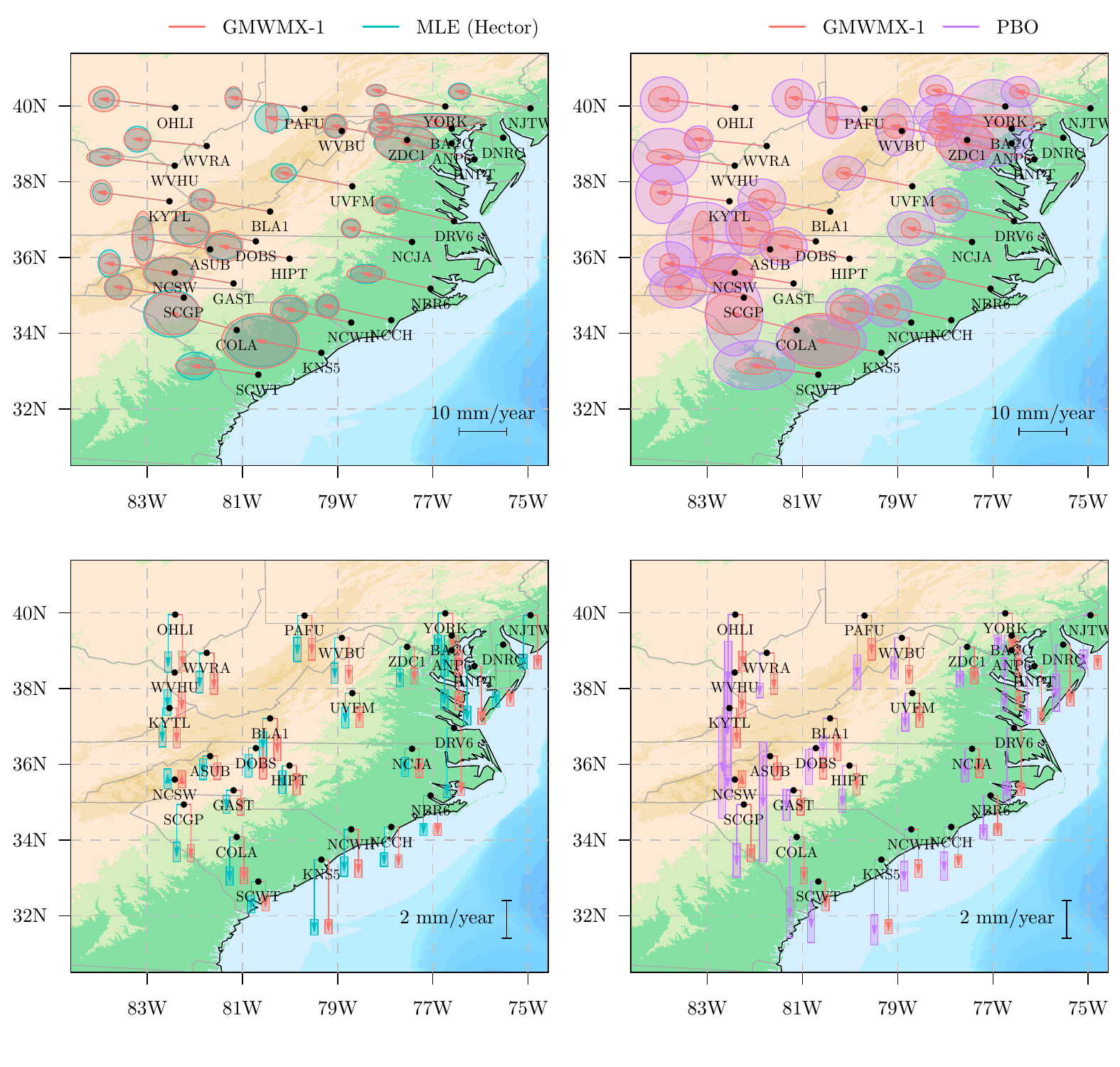}
	\caption{Estimated North-East velocity solutions and crustal uplift for $33$ GNSS receivers distributed over the East coast of the USA using i) the  GMWMX-1 ii) Hector software (MLE) iii) the PBO product.}
	\label{fig:mapAll}
\end{figure}

We apply our method to daily GNSS coordinate time series. We use measurements from $33$ continuously operating GNSS receivers distributed over the east coast of the USA. The daily position time series result from the processing released by the Pacific Northwest Geodetic Array at the Central Washington University (PANGA/CWU, \citealp{Herring2016,He2021}) computed within the International Terrestrial Reference Frame 2014 \citep{Altamimi2016}.

The analysis center PANGA/CWU computes the daily positions using the Precise Point Positioning method with the GIPSY software developed by NASA's Jet Propulsion Laboratory (JPL). The latter also provides the necessary satellite ephemerides, clock corrections, and wide-lane phase bias estimates~\citep{Herring2016}. The station positions were loosely constrained during the initial estimation and subsequently transformed into the International Terrestrial Reference Frame (ITRF2014) using only the translation and rotation~\citep{Altamimi2016}, but not scale, components of the JPL-provided Helmert transformations. Readers interested in the comprehensive discussion on the choice of the processing parameters can refer to~\cite{Herring2016} and~\cite{He2021}.

We use the resulting daily position solution time series to estimate the tectonic rate and the associated uncertainties with the GMWMX-1 and the MLE, as implemented in the Hector software~\citep{Bos2008}. For comparison purposes, we have also included the velocity solutions provided by the PBO~\citep{Herring2016}. The $33$ GNSS stations have at least $8$ years of continuous observations (see Figure \ref{fig:mapAll}). The same time range is carefully selected for each station in order to do a genuine comparison between the estimated tectonic rate with Hector, GMWMX-1 and the PBO solutions. The input data contains outliers. We employ the utility \texttt{removeoutliers} included in the Hector package since outlier rejection is beyond the scope of the current work.

For both GMWMX-1 and the MLE, we chose the functional model presented in~\eqref{eq:functionalmodel}, which includes a seasonal and half-seasonal component and multiple offsets. The offset times ($t_k$) are provided by PBO while we estimate the amplitudes ($g_k$). The stochastic model is a sum of a power-law and a white noise, also used to perform the simulations presented in the previous section.

To quantify the difference between the solutions from a statistical perspective, Figure \ref{fig:ratios} displays the range of rates and uncertainties, i.e. the ratio between the estimates and the associated uncertainties. Our solution (GMWMX-1) and the estimates with Hector (MLE) compare well within error. In terms of mean value, the ratio difference is $~ 0.6 \%$ (East), $~1.2 \%$ (North) and $~8.5 \%$ (Up). For the East component, the ratio is much higher than for the other ones suggesting that the uncertainty is small compared to the tectonic rate. Note that the Up component is known to contain $3$ times more noise than the horizontal coordinates \citep{Montillet2020Springer}. Correspondingly, the uncertainty is large resulting in a small ratio. Looking at the ratio difference between the PBO and GMWMX-1, the results are $~46 \%$ (East), $~50 \%$ (North) and $~23 \%$ (Up). This large difference is basically due to the uncertainty associated with the tectonic rate. Appendix~\ref{app.info} displays the results for GMWMX-2 which are very similar to the ones for GMWMX-1. Given that each GNSS station records observations for the three coordinates (East, North, Up) and that the mean size of each time series is approximately $10$ years, ranging between $8$ to $15$ years, the computing time for the GMWMX-1 for the whole GNSS network is around one minute, while in comparison, Hector's processing time is approximately $8$ hours. 
 
Figure \ref{fig:mapAll} displays the various solutions (i.e., GMWMX-1, Hector and PBO). Note that we have separated the arrows on the maps of the crustal uplift for the sake of clarity. The values are shown in Tables~\ref{tab:est_north}, \ref{tab:est_east} and \ref{tab:est_up} in the Appendix~\ref{app.info}. Overall, the solutions agree with the results published by \cite{PEROSANZ2019171} and \cite{Metivier2020}. The good agreement between Hector and GMWMX-1 can be seen visually for the East, North and Up components. They validate the results from the simulated time series and show the good agreement with Hector processing. The PBO solution is in line with the MLE and GMWMX-1 results for the amplitude of the tectonic rate and the crustal uplift. However, the uncertainties with this product are generally larger which is due to the difference between the methods. The GMWMX-1 and Hector are both jointly estimating a stochastic noise together with a geophysical model, whereas the PBO solution is based on a fast statistical approach. The method relies on a Kalman filter based on a first-order Gauss-Markov noise characteristic without any further analysis on the noise structure of the data \citep{Floyd2020}. The difference in the uncertainties is emphasized by the crustal uplift values. This result explains the ratio difference between the PBO solutions and the other methods in Figure~\ref{fig:ratios}. 

\section{Conclusions}
\label{sec:conclusions}

In this contribution, we propose a new method called the GMWMX to estimate the parameters of linear models with correlated residuals, which we apply to the analysis of GNSS daily position time series. The GMWMX allows a computationally efficient estimation of stochastic and functional (geophysical) models. Moreover, our approach is semi-parametric in the sense that the underlying distribution is left unspecified. Unlike the MLE, the GMWMX remains consistent and asymptotically normally distributed for all zero-mean probability distributions satisfying mild regularity conditions. Our approach is scalable in the sense that two estimators (GMWMX-1 and GMWMX-2) are proposed. The first estimator GMWMX-1 is particularly computationally efficient and over $1000$ times faster than the MLE for times series longer than $20$ years. However, this estimator comes at the price of marginally deteriorated statistical properties. The second estimator GMWMX-2 has an increased processing time but remains considerably faster than the MLE. Indeed, this estimator is approximately $40$ times faster to compute than the MLE for times series longer than 20 years. The GMWMX-2 is shown to be asymptotically efficient (and therefore asymptotically equivalent to the MLE) for the linear functional parameters. Moreover, this estimator corresponds to the (asymptotically) best unbiased estimator in the sense of \cite{hansen2022modern}. Both estimators are consistent and asymptotically normally distributed under arguably weak conditions (see \citealp{guerrier2013wavelet, guerrier2021robust} for details).

Our theoretical findings are validated considering different simulated scenarios. We consider several simulated cases based on different stochastic models. The impact of missing values is also investigated in the different simulated settings. Our results indicate that the GMWMX-1 is $400$ to $1200$ times faster than the MLE but comes at the price of a marginally inflated RMSE (around 5\% on average) compared to the MLE for the functional parameters. The GMWMX-2 is $20$ to $40$ times faster than the MLE but its statistical performance is indistinguishable from the MLE for the functional parameters (less than 0.01\% difference in terms of RMSE). Both the GMWMX-1 and the GMWMX-2 lead to comparable results to the MLE for the estimation of the stochastic parameters.

In order to support the simulation studies, we apply our algorithm to the analysis of real observations recorded from a network of $33$ GNSS stations located in the eastern part of the USA. These selected stations have registered at least $8$ years of data. Our results indicate that the use of the GMWMX-1 gives comparable results to the MLE with a widely assumed stochastic model (white noise summed with a power-law process). Overall, the results are nearly identical (difference of less than $1 \%$) between the MLE and GMWMX-1 when looking at the estimated tectonic rate and crustal uplift at each station, and the associated uncertainties. The clear advantage of the GMWMX-1 is the processing time which is approximately $600$ times lower than the one of the MLE with a marginal difference in terms of RMSE. Similar velocity estimates are obtained for the MLE and the GMWMX as well as for the stochastic parameters, highlighting the consistency of the two estimators. However, the associated uncertainties can vary up to $30 \%$ compared to the PBO solution. This large variation can be explained by the fast statistical approach used for the PBO solution which is based on an approximated stochastic noise model. 

The GMWMX allows to jointly estimate a functional and a stochastic noise model and produces accurately reliable uncertainties of the estimated parameters. It is a computationally efficient and scalable estimator based on simple statistical concepts and will be ideal to process large scale networks which include thousands of GNSS stations.

\section{Acknowledgments}
The data for PANGA/CWU processing center are available freely at \url{https://data.unavco.org/archive/gnss/products/position/}. In this study, we used, e.g., \url{YORK/YORK.pbo.igs14.pos} for the YORK station. The PBO velocity solutions are available at \url{https://data.unavco.org/archive/gnss/products/velocity/}. We choose the solutions \url{cwu.final_igs14.vel}.

D. A. Cucci and S.~Guerrier were supported by the SNSF Professorships Grant \#176843 and by the Innosuisse-Boomerang Grant \#37308.1 IP-ENG. L.~Voirol was supported by SNSF Grant \#182684. G. Kermarrec is supported by the Deutsche Forschungsgemeinschaft under the project KE2453/2-1.

We thank to Roberto Molinari, Maria-Pia Victoria-Feser, Haotian Xu and Yuming Zhang for their valuable inputs and suggestions throughout the development of this work as well as for their comments on the final draft.

The GMWMX method is available in the \texttt{R} package \texttt{gmwmx}, which is available on Github \url{https://github.com/SMAC-Group/gmwmx}.

\bibliographystyle{chicago}
\bibliography{refs}

\newpage
\appendix

\begin{center}
  \LARGE \textsc{Appendices}  
\end{center}

\setcounter{equation}{0}
\renewcommand{\theequation}{\thesection.\arabic{equation}}

\section{Mathematical Derivation} 
\label{app:derivation:1}

Here, we focus on the derivation of~\eqref{eq:pos:def:mat}. By  comparing~\eqref{eq:step1:limit} with~\eqref{eq:asym:mle} we have
\begin{equation*}
\begin{aligned}
   \lim_{n \to \infty} \;  &\var\left\{\sqrt{n} \left( \widetilde{\bm{\x}} - \bm{\x}_0 \right)\right\} - \var\left\{\sqrt{n} \left( \widehat{\bm{\x}} - \bm{\x}_0 \right)\right\} = 
    n\left(\mathbf{A}\trans   \mathbf{A} \right)^{-1}  \mathbf{A}\trans   \bm{\Sigma}(\bm{\gamma}_0) \mathbf{A} \left(\mathbf{A}\trans  \mathbf{A} \right)^{-1} - n\left(\mathbf{A}\trans  \bm{\Sigma}(\bm{\gamma}_0)^{-1} \mathbf{A} \right)^{-1} \\
    &= n \left(\mathbf{A}\trans   \mathbf{A} \right)^{-1} \mathbf{A}\trans \bm{\Sigma}(\bm{\gamma}_0)^{1/2} \left\{\I - \bm{\Sigma}(\bm{\gamma}_0)^{-1/2} \A \left(\mathbf{A}\trans  \bm{\Sigma}(\bm{\gamma}_0)^{-1} \mathbf{A} \right)^{-1} \A\trans \bm{\Sigma}(\bm{\gamma}_0)^{-1/2}\right\} \bm{\Sigma}(\bm{\gamma}_0)^{1/2} \mathbf{A} \left(\mathbf{A}\trans   \mathbf{A} \right)^{-1}.
\end{aligned}
\end{equation*}
Then, by defining $\B = \bm{\Sigma}(\bm{\gamma}_0)^{-1/2} \A \left(\mathbf{A}\trans  \bm{\Sigma}(\bm{\gamma}_0)^{-1} \mathbf{A} \right)^{-1} \A\trans \bm{\Sigma}(\bm{\gamma}_0)^{-1/2}$ and noticing that $\B$ is idempotent, we obtain
\begin{equation*}
    \begin{aligned}
    &n \left(\mathbf{A}\trans   \mathbf{A} \right)^{-1} \mathbf{A}\trans \bm{\Sigma}(\bm{\gamma}_0)^{1/2} \left\{\I - \bm{\Sigma}(\bm{\gamma}_0)^{-1/2} \A \left(\mathbf{A}\trans  \bm{\Sigma}(\bm{\gamma}_0)^{-1} \mathbf{A} \right)^{-1} \A\trans \bm{\Sigma}(\bm{\gamma}_0)^{-1/2}\right\} \bm{\Sigma}(\bm{\gamma}_0)^{1/2} \mathbf{A} \left(\mathbf{A}\trans   \mathbf{A} \right)^{-1}\\
    =& n \left(\mathbf{A}\trans   \mathbf{A} \right)^{-1} \mathbf{A}\trans \bm{\Sigma}(\bm{\gamma}_0)^{1/2} \left(\I -\B\right) \left(\I -\B\right)\trans \bm{\Sigma}(\bm{\gamma}_0)^{1/2} \mathbf{A} \left(\mathbf{A}\trans   \mathbf{A} \right)^{-1}\\
    =& n \left\{\left(\mathbf{A}\trans   \mathbf{A} \right)^{-1} \mathbf{A}\trans \bm{\Sigma}(\bm{\gamma}_0)^{1/2} \left(\I -\B\right)\right\} \left\{\left(\mathbf{A}\trans   \mathbf{A} \right)^{-1} \mathbf{A}\trans \bm{\Sigma}(\bm{\gamma}_0)^{1/2} \left(\I -\B\right)\right\}\trans.
    \end{aligned}
\end{equation*}
Thus, we obtain
\begin{equation*}
   \lim_{n \to \infty} \;   \var\left\{\sqrt{n} \left( \widetilde{\bm{\x}} - \bm{\x}_0 \right)\right\} - \var\left\{\sqrt{n} \left( \widehat{\bm{\x}} - \bm{\x}_0 \right)\right\} > \mathbf{0},
\end{equation*}
for $\bm{\Sigma}(\bm{\gamma}_0) \, \cancel{\propto}\,\, \mathbf{I}$.

\section{Further Discussion on the Generalized Method of Wavelet Moments}
\label{app:gmwm}

As mentioned in Section~\ref{sec:stat:part}, the GMWM estimator based on an estimator of $\x_0$, say $\x$, is defined as follows:
\begin{equation*}
    \wt{\bm{\gamma}}\left(\x\right) = \underset{\bm{\gamma} \in \bm{\Gamma}}{\argmin}\; \left\{\wh{\bm{\nu}}\left({\x}\right) - \bm{\nu}(\bm{\gamma})\right\}\trans\bm{\Omega} \left\{\wh{\bm{\nu}}\left({\x}\right) - \bm{\nu}(\bm{\gamma})\right\},
\end{equation*}
where $\bm{\nu}(\bm{\gamma})$ is the WV vector implied by the model and $\wh{\bm{\nu}}\left({\x}\right)$ is the estimated Haar WV computed on ${\bm{\varepsilon}}\left({\x}\right)$. To define these quantities, we let
\begin{equation*}
	W_{j,t}(\x) = \sum_{l = 0}^{L_j - 1} h_{j,l} {\bm{\varepsilon}}\left({\x}\right)_{t - l} \, ,
\end{equation*}
denote the wavelet coefficients associated to the (Haar) Maximal Overlap Discrete Wavelet Transform (MODWT) wavelet decomposition of the time series (see e.g.,,,,,,,,,, \citealp{percival2000wavelet}), where $(h_{j,t})$ is a known wavelet filter of length $L_j$ at scale $\tau_j = 2^j$, for $j = 1, \hdots, J$ and $J < \log_2(n)$. Based on this quantity, for $j = 1, \dots, J$ the empirical WV at scale $j$ is defined as
\begin{equation*}
\nu_{j}^2(\x) = \var\{W_{j,t}(\x)\},
\end{equation*}
which corresponds to the variance of the wavelet coefficients. The vector $\wh{\bm{\nu}}\left({\x}\right)$ can then be expressed as $\bm{\nu}(\x) = [\nu_j^2(\x)]_{j = 1, \hdots, J}$. Several estimators have been proposed for the WV, we consider here the MODWT WV estimator proposed in \citet{percival1995estimation}, which enjoys from desirable statistical properties. This estimator is simply defined as
\begin{equation}
    \wh{\nu}_{j}^2(\x) = \frac{1}{M_{j}} \sum_{t=1}^{M_j} W_{j, t}^2(\x) \,\,,
    \label{eq:standard_wvest}
\end{equation}
where $M_j$ is the length of the wavelet coefficient process $(W_{j, t})$ at scale $\tau_j$. We define $\wh{\bm{\nu}}\left({\x}\right) = [\wh{\nu}_{j}^2(\x)]_{j = 1, \hdots, J}$. A detailed introduction of the WV can be found in \cite{percival2000wavelet} and the references therein. Moreover, the theoretical properties of this estimator were further studied in for example \cite{serroukh2000statistical} and \cite{guerrier2021robust} in which the conditions for its asymptotic properties are provided under different frameworks (such as those considered in this contribution).

In order to make the link between the WV and an assumed {stationary} (or intrinsically stationary) parametric model explicit, we have the following relation between the WV and the Spectral Density Function (SDF):
\begin{equation}
	\nu_{j}^2 (\bg) = \int_{-1/2}^{1/2} |H_j(f)|^2 S(f) df \, ,
	\label{eq:wv:theo}
\end{equation}
with $S(f)$ denoting the theoretical SDF and $H_j(f)$ being the Fourier transform of the wavelet filters $(h_{j,t})$. In practice, computing ~\eqref{eq:wv:theo} is often difficult but the results of \cite{zhang2008allan} provide the following result:
\begin{equation}
    \nu_{j}^2 (\bg) = \frac{\bm{\Sigma}(\bg)_{1,1}}{2^{(2j - 1)}}  \left[
    2^{j - 1} \left(1 - \bm{\Sigma}(\bg)_{1,2^{(j-1)}}\right) + \sum_{i = 1}^{2^j - 1} i \left\{2 \bm{\Sigma}(\bg)_{1,2^(j-1) - i} - \bm{\Sigma}(\bg)_{1,i} - \bm{\Sigma}(\bg)_{1,2^j - i}\right\} 
    \right],
    \label{eq:theo:wv}
\end{equation}
which allows to simply compute the theoretical WV of nearly any intrinsically stationary parametric stochastic process.

\section{Additional simulation study}
\label{app:matern}

In this section we present the results of an additional simulation study in which the power-law stochastic process employed in Section~\ref{sec:simu} has been replaced with a Mat\`ern model.
\begin{figure}[H]
	\centering
	\includegraphics[]{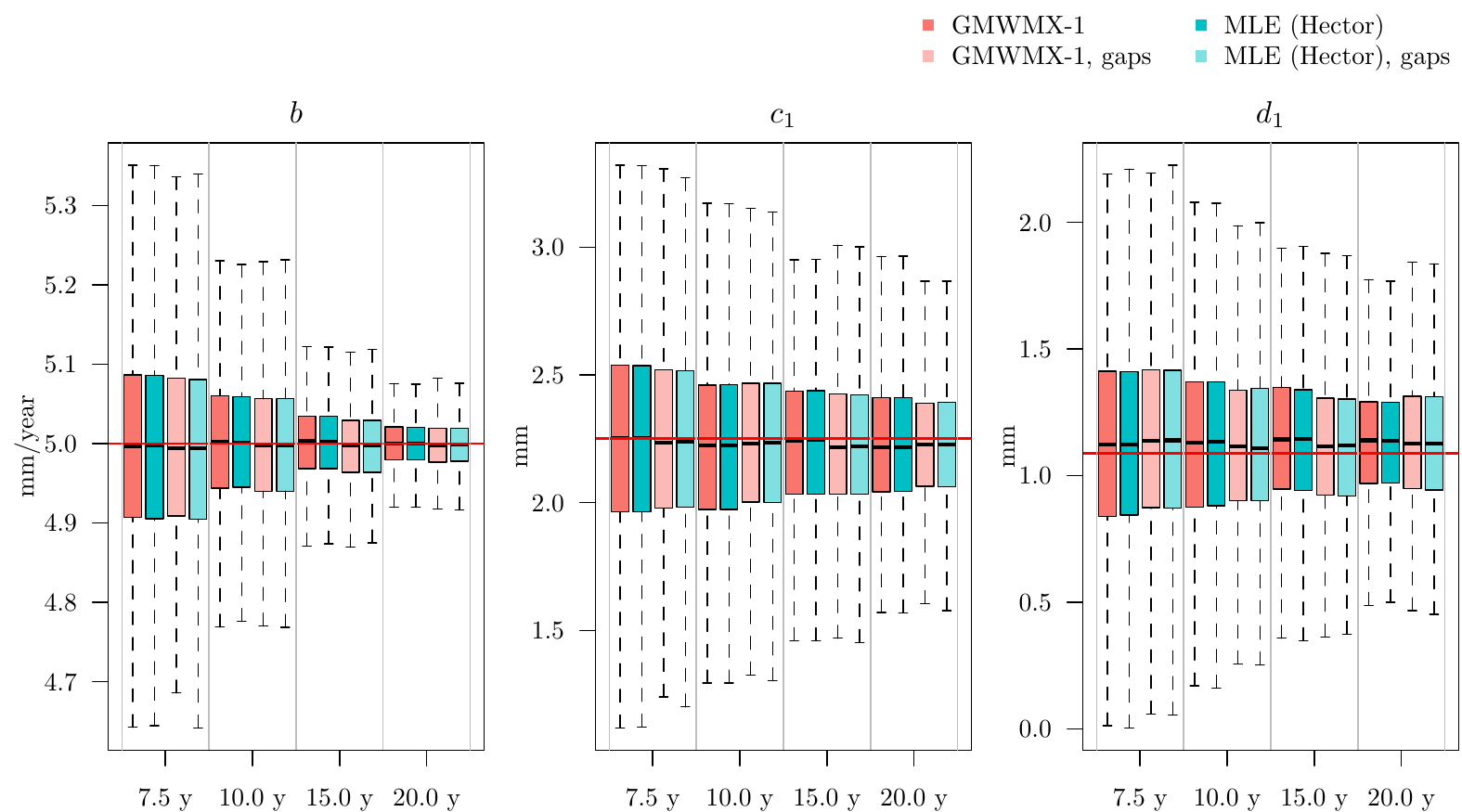}
	\caption{Boxplots of the estimated deterministic parameters with method GMWMX-1 and the MLE for the scenario presenting all the data and no offsets and the realistic scenario. The red line indicate the true value of the estimated parameter.}
	\label{fig:detparams_addi_simu}
\end{figure}

\begin{figure}[H]
	\centering
    \includegraphics[]{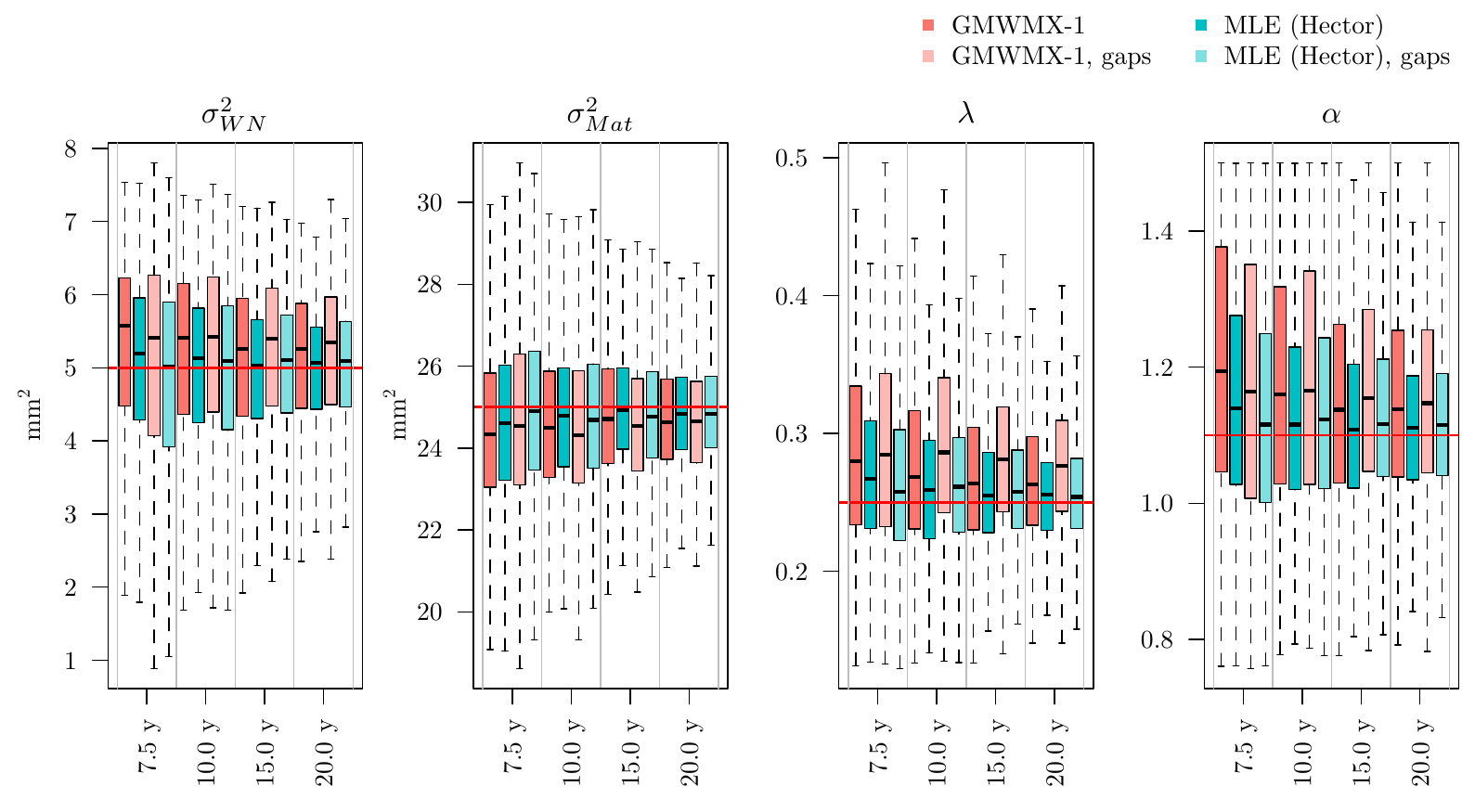}
	\caption{Boxplots of the estimated stochastic parameters with method GMWMX-1 and the MLE for the scenario presenting all the data and no offsets and the realistic scenario. The red line indicate the true value of the estimated parameter. }
	\label{fig:stochparams_addi_simu}
\end{figure}

In Figure \ref{fig:detparams_addi_simu}, we can see that the results are similar for the estimation of the geophysical parameters in terms of mean and uncertainties. However, the estimation of the stochastic noise parameters (see Figure \ref{fig:stochparams_addi_simu}) is slightly worst than the power-law noise model. This result is valid for both GMWM and Hector. The reason is not fully known. Because the Mat\`ern process is a diffusion process, stochastic properties vary differently than a stationary process.

\setcounter{equation}{0}
\section{Confidence Intervals with the GMWMX}
\label{ap:ci}

We define the confidence interval used in the Section \ref{sec:study}. A $1-\alpha$ confidence interval for a parameter $\theta$ denoted $C_{n}=(f_{1}, f_{2})$ is the interval with random endpoints $f_{1}$ and $f_{2}$ where $f_{1}=f_{1}\left(X_{1}, \ldots, X_{n}\right)$ and $f_{2}=f_{2}\left(X_{1}, \ldots, X_{n}\right)$ are functions of the data such that
\begin{equation}
\Pr \left[ f_{1}(X_{1}, \ldots, X_{n}) \leq \theta \leq f_{2}(X_{1}, \ldots, X_{n}) \right] =1-\alpha.
\end{equation}
We call $1-\alpha$ the nominal coverage of the confidence interval.

An interesting property of the proposed estimators $\wt{\bt}_1$ and $\wt{\bt}_2$, which results from their consistency as well as the asymptotic distribution of $\wt{\x}^{(1)}$ and $\wt{\x}^{(2)}$, is that valid confidence intervals can be constructed for $\x_0$. For $i = \{1, \ldots, p\}$, we let
\begin{equation}
    \wt{\sigma}_{i,j}^2 = \left\{
	\begin{array}{ll}
		\left[\left(\mathbf{A}\trans   \mathbf{A} \right)^{-1}  \mathbf{A}\trans   \bm{\Sigma}\left(\widetilde{\bm{\gamma}}^{(1)}\right) \mathbf{A} \left(\mathbf{A}\trans  \mathbf{A} \right)^{-1}\right]_{i,i}  & \mbox{if } j = 1, \\
		\left[\left\{\mathbf{A}\trans  \left[\bm{\Sigma}\left(\widetilde{\bm{\gamma}}^{(2)}\right)\right]^{-1} \mathbf{A} \right\}^{-1}\right]_{i,i}  & \mbox{if } j = 2.
	\end{array}
\right.
    \label{eq:confint1}
\end{equation}

Using results of the asymptotic normality of $\wt{\x}^{(1)}$ and $\wt{\x}^{(2)}$ and Lemma 2.11 in \cite{van2000asymptotic} it can be shown that
\begin{equation*}
    \lim_{n \to \infty} \; \Pr\left[\left(\x_0\right)_i \in \left\{   \big(\wt{\x}^{(j)} \big)_i \pm z_{1-\alpha/2} \wt{\sigma}_{i,j} \right\} \right] = 1-\alpha.
    \label{eq:confint2}
\end{equation*}
where $z_{1-\alpha/2}$ is the $1-\alpha/2$ quantile of the standard normal distribution ($\Phi^{-1}(1-\alpha/2)$ where $\Phi$ denotes the cumulative distribution function of the standard Normal distribution).
Moreover if we assume that for $j \in \{1,2\}$ and for all $i \in \{1, \ldots, p\}$ the distribution of 
\begin{equation*}
    \frac{\sqrt{n} \left( \wt{\x}^{(j)} - \x_0 \right)_{i}}{\wt{\sigma}_{i,j}},
\end{equation*}
admits an Edgeworth expansion, a requirement that is generally satisfied under suitable regularity conditions (usually moment and smoothness conditions, see \citealp{hall2013bootstrap} and the references therein) then we have for all $\alpha \in (0, 0.5)$ and for all $i \in \{1, \ldots, p\}$ that
\begin{equation*}
    \Pr\left[\left(\x_0\right)_i \in \left\{   \big(\wt{\x}^{(j)} \big)_i \pm z_{1-\alpha/2} \wt{\sigma}_{i,j} \right\} \right] = 1-\alpha + \mathcal{O}\left(n^{-1/2}\right),
    \label{eq:confint3}
\end{equation*}
demonstrating the asymptotic validity of confidence intervals constructed with the proposed approach.

\section{Additional Information}
\label{app.info}

\begin{figure}[!htb]
	\centering
	\includegraphics[]{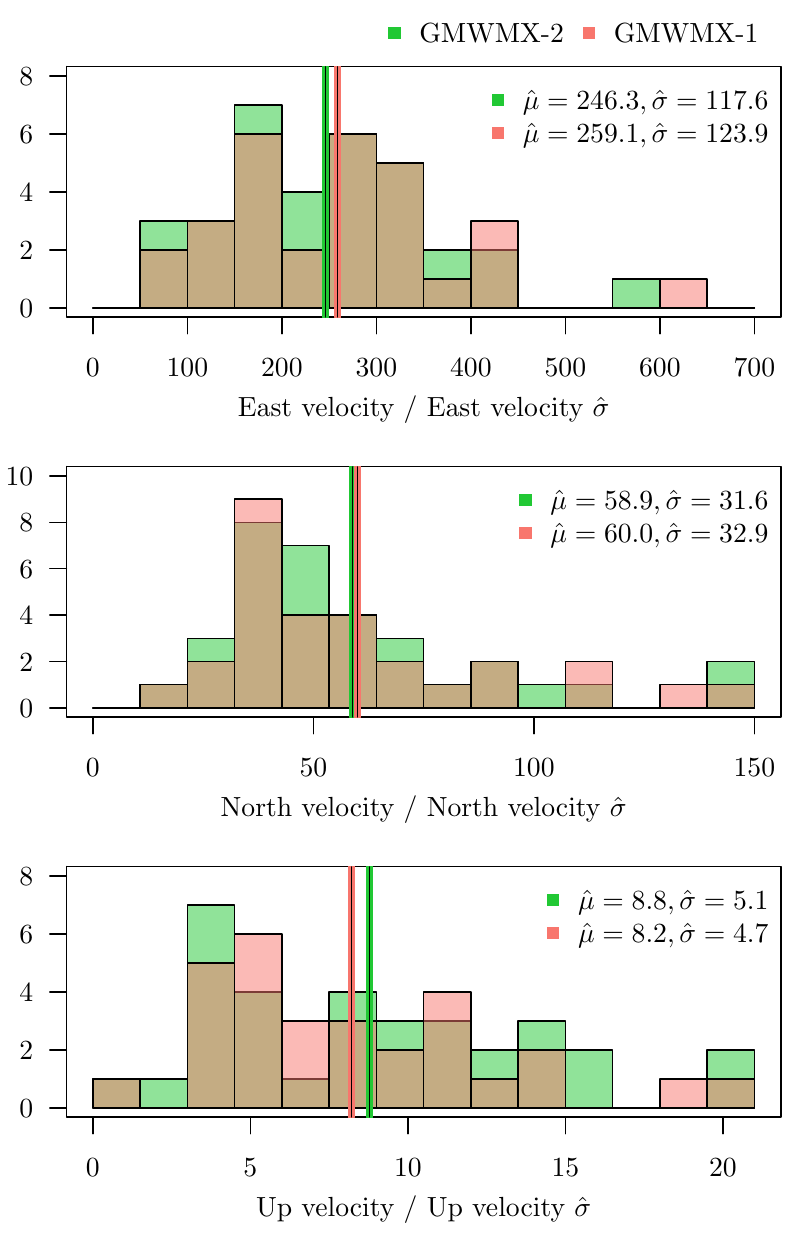}
		\caption{Ratios between estimated North-East velocities and crustal uplift divided by their respective estimated standard deviation for the GMWMX-2 and the GMWMX-1 for $33$ GNSS receivers distributed over the east coast of the USA.}
	\label{fig:ratios_gmwmx_2}
\end{figure}

\begin{table}
    \centering
    \includegraphics[scale=.94]{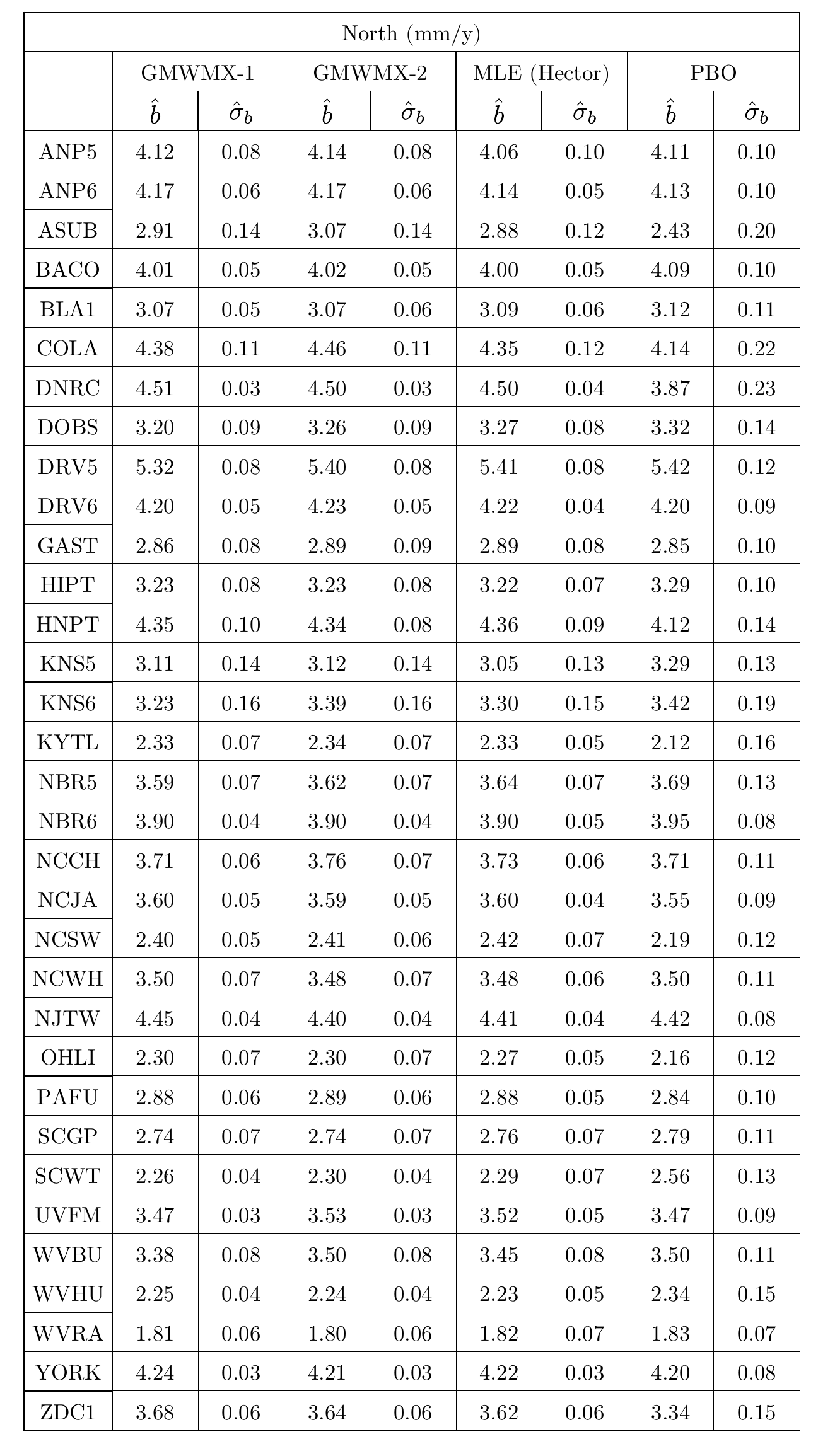}
    \caption{Estimated North velocities for the 33 stations considered in the case study using the GMWMX-1, GMWMX-2, the MLE as implemented in Hector and PBO velocity estimates. The estimated velocity for each method is denoted by $\hat{\mu}$ and $\hat{\sigma}_b$ denotes its estimated uncertainty (standard error). }
    \label{tab:est_north}
\end{table}

\begin{table}
    \centering
    \includegraphics[scale=.94]{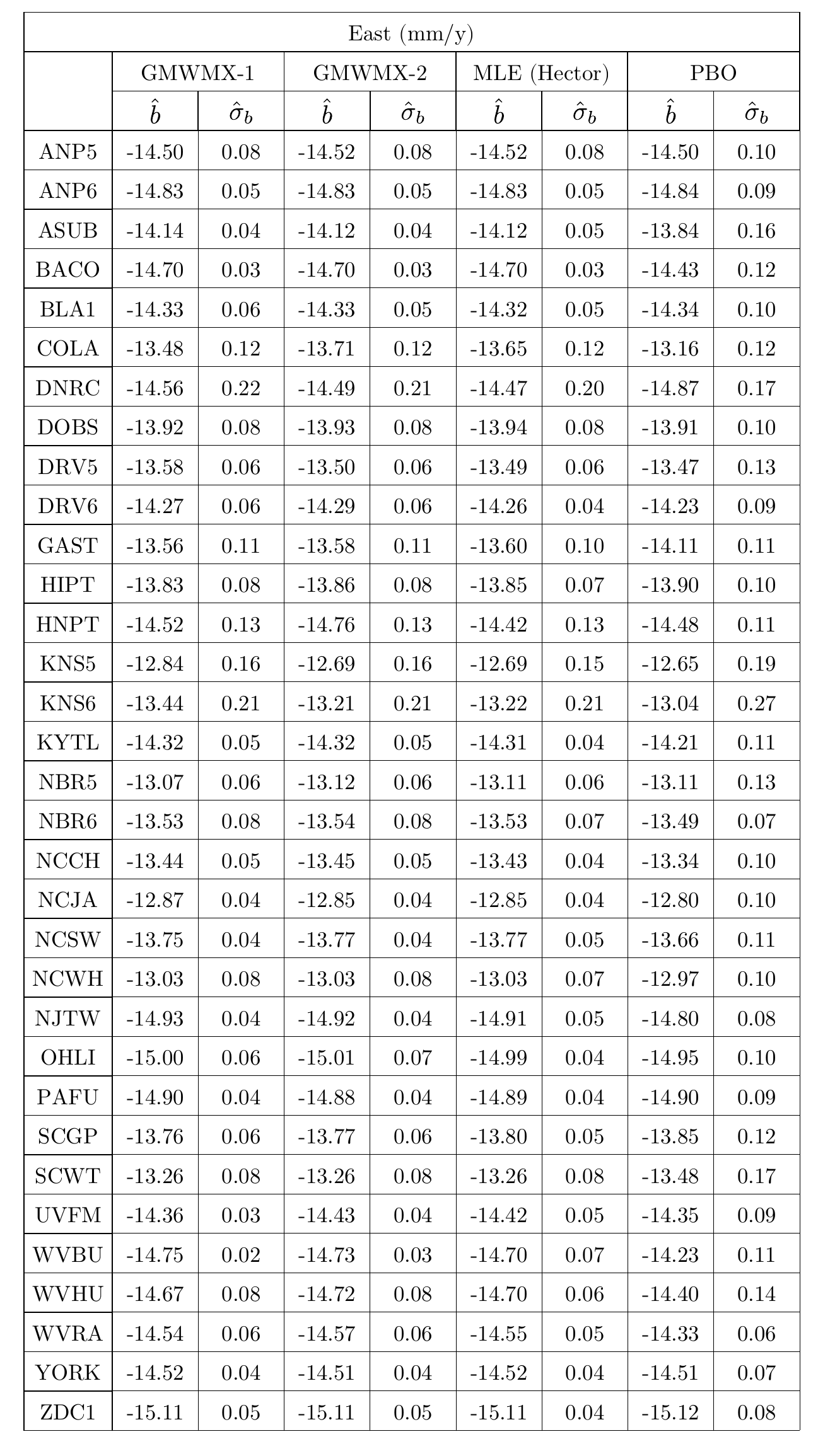}
    \caption{Estimated East velocities for the 33 stations considered in the case study using the GMWMX-1, GMWMX-2, the MLE implemented in Hector and PBO velocity estimates.  The estimated velocity for each method is denoted by $\hat{\mu}$ and $\hat{\sigma}_b$ denotes its estimated uncertainty (standard error). }
    \label{tab:est_east}
\end{table}

\begin{table}
    \centering
    \includegraphics[scale=.94]{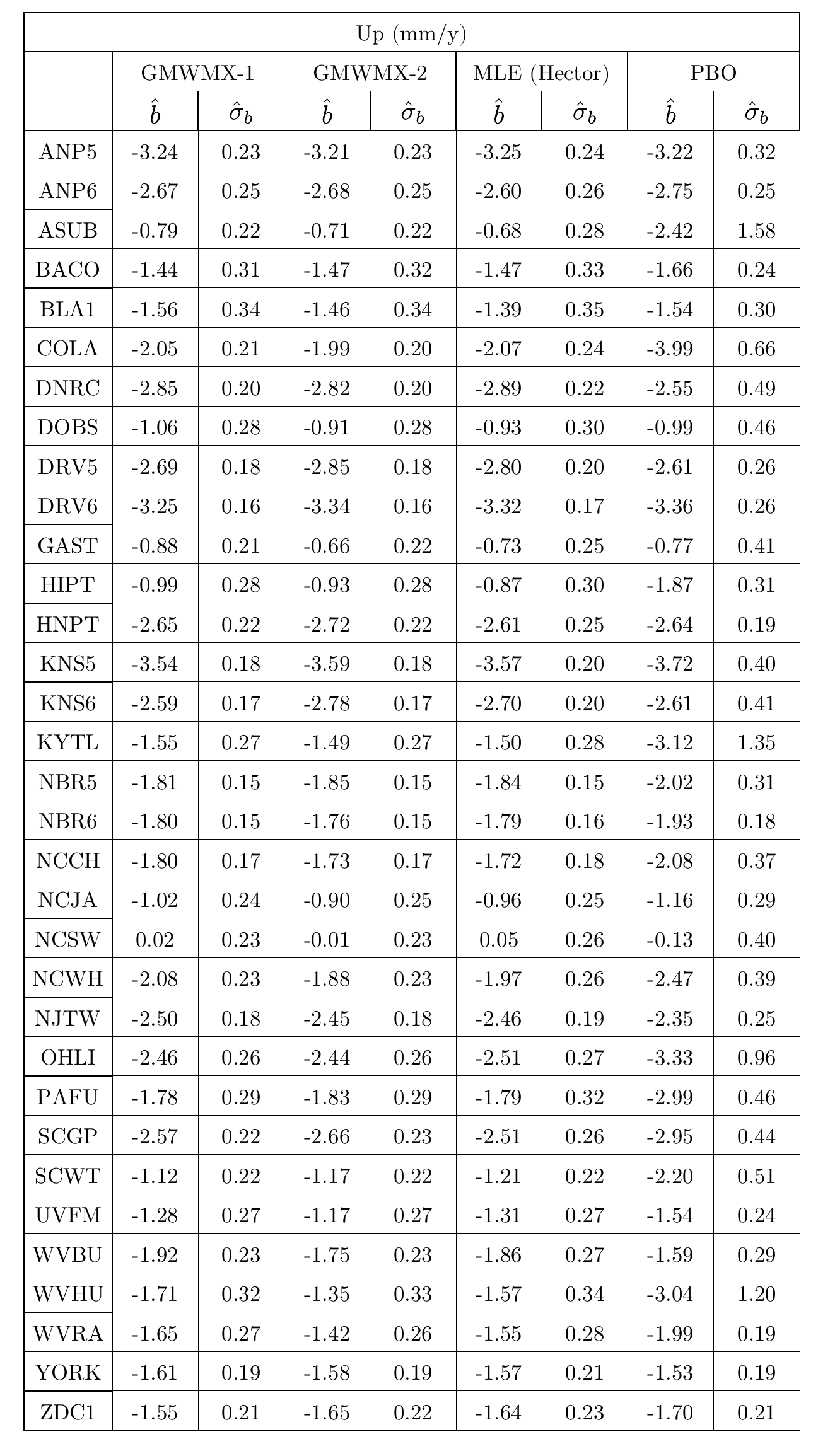}
    \caption{Estimated crustal uplift for the 33 stations considered in the case study using the GMWMX-1, GMWMX-2, the MLE implemented in Hector and PBO velocity estimates. The estimated velocity for each method is denoted by $\hat{\mu}$ and $\hat{\sigma}_b$ denotes its estimated uncertainty (standard error). }
    \label{tab:est_up}
\end{table}

\end{document}